\documentclass[aps,prb,superscriptaddress,twocolumn]{revtex4-1}

\usepackage{amsmath}
\usepackage{amssymb}
\usepackage{mathtools}
\usepackage{graphicx}
\usepackage[usenames,dvipsnames]{color}
\usepackage{bm}
\usepackage{hyperref}
\usepackage{url}
\usepackage[utf8]{inputenc}
\usepackage{subfigure}
\usepackage{slashed,bbm}
\usepackage{graphics,psfrag,epsfig}
\usepackage{dsfont}
\usepackage{setspace}
\usepackage{soul}

\begin{document}

\definecolor{debianred}{rgb}{0.84, 0.04, 0.33}
\definecolor{checkcolor}{rgb}{0.44, 0.04, 0.83}

\newcommand{\beq}{\begin{equation}}
\newcommand{\eeq}{\end{equation}}
\newcommand{\beqa}{\begin{eqnarray}}
\newcommand{\eeqa}{\end{eqnarray}}
\newcommand{\note}[1]{{\color{red} [#1]}}
\newcommand{\bra}[1]{\ensuremath{\langle#1|}}
\newcommand{\ket}[1]{\ensuremath{|#1\rangle}}
\newcommand{\bracket}[2]{\ensuremath{\langle#1|#2\rangle}}
\renewcommand{\vec}[1]{\bm{#1}}
\newcommand{\dagga}{{\phantom{\dagger}}}

\newcommand{\txi}{\tilde{\xi}}
\newcommand{\teta}{\tilde{\eta}}

\newcommand{\nn}{\nonumber } 

\newcommand{\LC}[1]{\textcolor{blue}{#1}}
\newcommand{\MS}[1]{\textcolor{debianred}{#1}}
\newcommand{\checkit}[1]{\textcolor{checkcolor}{#1}}

\allowdisplaybreaks

\title{\mbox{\hspace{-0.6cm} Competing phases of interacting electrons on triangular lattices in moir\'e heterostructures}}

\author{Laura Classen}
\affiliation{Physics Department, Brookhaven National Laboratory, Building 510A, Upton, New York 11973, USA}

\author{Carsten Honerkamp}
\affiliation{Institut f\"ur Theoretische Festk\"orperphysik, RWTH Aachen University, and JARA Fundamentals of Future Information Technology, Germany}

\author{Michael M. Scherer}
\affiliation{Institute for Theoretical Physics, University of Cologne, 50937 Cologne, Germany}

\date{\today}

\begin{abstract}
We study the quantum many-body instabilities of interacting electrons with SU(2)$\times$SU(2) symmetry in spin and orbital degrees of freedom on the triangular lattice near van-Hove filling. Our work is motivated by effective models for the flat bands in hexagonal moir\'e heterostructures like twisted bilayer boron nitride and trilayer graphene-boron nitride systems.
We consider an extended Hubbard model including onsite Hubbard and Hund's couplings, as well as nearest-neighbor exchange interactions and analyze the different ordering tendencies with the help of an unbiased functional renormalization group approach.
We find three classes of instabilities controlled by the filling and bare interactions.
For a nested Fermi surface at van-Hove filling, Hund-like couplings induce a weak instability towards spin or orbital density wave phases. An SU(4) exchange interaction moves the system towards a Chern insulator, which is robust with respect to perturbations from Hund-like interactions or deviations from perfect nesting. Further, in an extended range of fillings and interactions, we find topological $d\pm id$ and (spin-singlet)-(orbital-singlet) $f$-wave superconductivity.
\end{abstract}

\maketitle

%
Correlated insulating and superconducting behavior have recently been discovered in twisted bilayer graphene (TBG)~\cite{cao2018correlated,cao2018unconventional,yankowitz2018tuning} triggering ample excitement due to their potential to shed new light on the problem of unconventional superconductivity.
More generally, experiments with moir\'e superlattices of two-dimensional van-der-Waals heterostructures -- further including, for example, hexagonal boron nitride (hBN) layers~\cite{chen2018gate} -- are established as an experimental platform for studies of correlated electron physics.
These systems allow for a high degree of control, e.g., in the regulation of the twist angle, a low level of disorder and gate-tunable effective bandwidths or filling factors.
Indeed, signatures of tunable insulating and superconducting states have been reported in trilayer graphene/hBN heterostructures~\cite{chen2018gate,chen2019signatures}.
Further, it has been suggested that a wider class of ``magic-angle'' systems 
can be realized in present cold-atom setups \cite{fu2018magic}.
The appearance of strong correlations in these systems is usually ascribed to the emergence of low-lying flat bands and an increased density of states amplifying the impact of electronic interactions\cite{doi:10.1021/nl902948m,Bistritzer12233,PhysRevB.86.155449,li2010observation,PhysRevLett.109.196802,doi:10.1021/acs.nanolett.6b01906}.
However, in conjunction with the experimental findings, the nature of correlated states in moir\'e superlattices has yet to be identified and described by appropriate models and methods.
One way, recently pursued, is the construction of effective multi-orbital tight-binding models for the nearly flat bands based on Wannier states~\cite{PhysRevX.8.031087,PhysRevX.8.031088,PhysRevX.8.031089}. 
While this in itself is a non-trivial task, some universal aspects are shared by all models, including the focus on the emergent superlattice, the presence of several orbitals inherited from the valleys of the original bands and sizable further-neighbor interactions.
Hence, a basic understanding of the correlated behavior may be developed by analyzing phenomenological models that capture the qualitative features of moir\'e flatbands~\cite{PhysRevLett.121.087001,PhysRevB.98.045103,PhysRevB.98.075154,PhysRevB.98.241407,tang2018spin,PhysRevX.8.041041,lin2019chiral,PhysRevB.98.235158,huang2018antiferromagnetically,PhysRevB.98.085436,PhysRevB.97.235453,roy2018unconventional,PhysRevLett.121.217001,da2019magic}.
In addition, this approach allows one to study multi-orbital effects in hexagonal systems in general. This opens the possibility to new types of interactions and correlations. For example,  spin-singlet(triplet) pairing is not bound to even(odd)-parity because the anti-symmetry can be guaranteed through the orbital degrees of freedom.

\begin{figure}[t!]
\includegraphics[width=0.9\columnwidth]{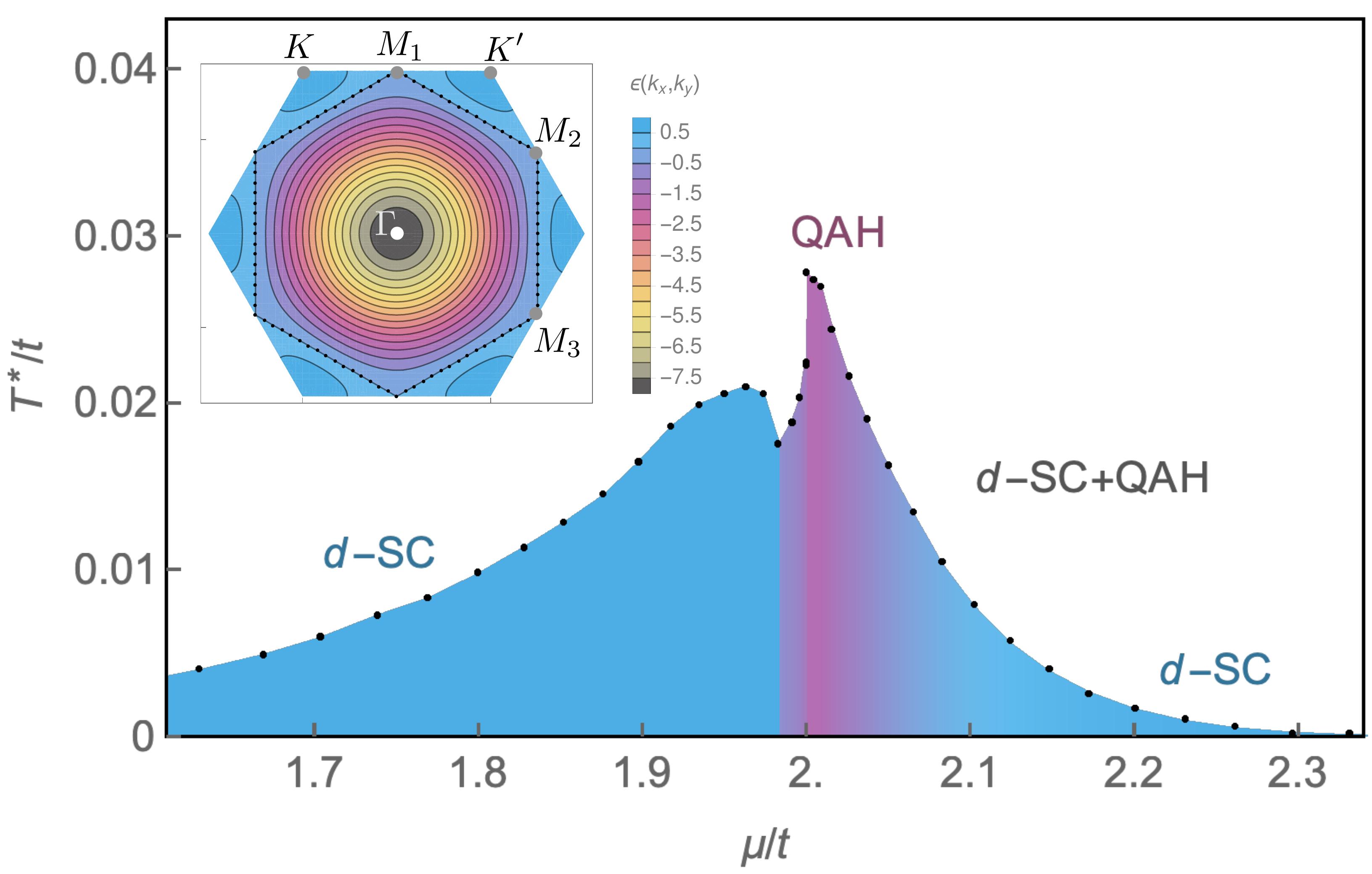}
\caption{Tentative phase diagram showing the quantum anomalous Hall (QAH) instability near van-Hove filling $\mu=2t$ flanked by the $d$-wave superconducting states based on our functional renormalization group calculations.
We plot the instability temperature $T^\ast$ for initial couplings $U=3t, J=0.4t, V_h=0.1t$, and variable chemical potential $\mu$. Unspecified initial couplings are zero. Inset: Brillouin zone of the triangular lattice with Fermi level at the van-Hove singularity, Eq.~\eqref{eq:tightbinding}. Black dots illustrate our discretization of the Fermi surface.}
\label{fig:regions}
\end{figure}

Regarding the role of interactions, no consensus has been reached on whether the experimentally observed insulating states in TBG or trilayer graphene/hBN are due to Mott physics in the strong coupling limit or due to interaction-driven symmetry-breaking, e.g., based on Fermi-surface nesting close to van-Hove singularities~\cite{li2010observation,PhysRevLett.109.196802,doi:10.1021/acs.nanolett.6b01906,PhysRevX.8.041041,lin2019chiral}.
For the lattice structure of the effective model, triangular and honeycomb geometries have been discussed. While for TBG most recent suggestions favor a honeycomb superlattice~\cite{PhysRevX.8.031087,PhysRevX.8.031088,PhysRevX.8.031089}, moir\'e structures with hBN are argued to support a triangular superlattice, e.g., trilayer graphene/hBN~\cite{chen2018gate,PhysRevX.8.031089} and twisted bilayer hBN~\cite{2018arXiv181208097X}.
In these systems, the flat bands are suggested to be well isolated~\cite{PhysRevLett.122.016401,chen2018gate,2018arXiv181208097X} so that application of an effective model  for the flat bands seems appropriate.

In this work, we consider a phenomenological model of electrons on a triangular lattice with two spin and two orbital states as well as onsite and nearest-neighbor interactions. We follow earlier arguments that the combined spin and orbital degrees of freedom form an approximate SU(4) flavor symmetry~\cite{PhysRevLett.121.087001,PhysRevB.98.075154,PhysRevB.98.045103,PhysRevB.98.241407} and perturb it by Hund-like interactions down to SU(2)$\times$SU(2). We assume that the experimental observations correspond to fillings close to a van-Hove singularity and a (near-)nested Fermi surface and determine potential ordered states induced by moderate interactions. In this situation, electronic correlations in several channels are important, which is why we approach the problem with the help of the fermion functional renormalization group (fRG) method~\cite{RevModPhys.84.299,doi:10.1080/00018732.2013.862020}.

Several other works have analyzed multi-orbital models on hexagonal lattices close to van-Hove filling with a nested Fermi surface and repulsive interactions. We extend different aspects of these works: 
in contrast to random phase approximation (RPA) approaches~\cite{PhysRevLett.121.217001}, we take into account the essential coupling between particle-particle and particle-hole channels. 
While such competition between the different channels is included in parquet RG studies~\cite{PhysRevX.8.041041,lin2019chiral}, our calculations are not restricted to a small number of patches at the van-Hove singularity, but resolve the momentum dependence of the entire Fermi surface. 
This allows us to draw a direct connection to microscopic models and upon inclusion of  model parameters to provide quantitative estimates on doping ranges and energy scales.
Further, the fRG equations are well defined rooting in an exact flow equation~\cite{Wetterich:1992yh} and include not only the leading $\ln^2$-diverging channels, but also sub-leading $\ln$-divergent contributions. 
We extend previous fRG studies~\cite{PhysRevB.98.241407,tang2018spin,2018arXiv181208097X} by inclusion of other interaction types and a more comprehensive exploration of the induced ordering tendencies. The exploitation of SU(2)$\times$SU(2) symmetry facilitates high Fermi-surface resolution, which is important near van-Hove singularities.

We find that the interplay between (approximate) nesting, a large density of states and interaction terms results in three classes of strongly-growing correlations: (1) spin/orbital density wave states, (2) $d\pm id$ and $f$-wave superconductivity and (3) a Haldane-like loop-current phase. 
The density waves are generated right around perfect nesting and involve the different SU(2)$\times$SU(2) (pseudo-)spins.
It depends on the SU(4)-symmetry-breaking interactions if a density wave in the separate SU(2) spin or orbital sector, or in the combined SU(2)$\times$SU(2) sector develops. 
Due to the larger number of fermion flavors and available nesting vectors in the hexagonal geometry, however, the density wave instabilities are fragile.
Instead, superconducting and loop-current phases are more robust, see also Ref.~\onlinecite{lin2019chiral}.
Which pairing instability emerges depends on the fermiology: for a closed Fermi surface we find topological $d\pm id$ superconductivity\cite{PhysRevLett.121.087001,lin2019chiral}, for an open Fermi surface pairing correlations in either $d$- or $f$-channel grow strong. The loop-current state also develops around the perfectly nested situation as soon as small nearest-neighbor SU(4) exchange interactions are present. This quantum-anomalous-Hall state breaks time-reversal symmetry and has a non-zero Chern number together with a fully gapped spectrum\cite{PhysRevB.93.115107,lin2019chiral}.

\textit{Model and parameters.} 
%
We follow early phenomenological approaches \cite{PhysRevLett.121.087001,PhysRevB.98.075154,lin2019chiral} and model
the superlattice of twisted hexagonal multi-layered systems 
by a two-dimensional triangular lattice populated by electrons with SU(2) spin and orbital degrees of freedom.
In particular for twisted hBN, this seems to be an appropriate starting point, where the dispersion close to the Fermi level is well-matched by a two-orbital tight-binding model with nearest-neighbor hopping on the superlattice with hopping amplitude $t(\alpha)$ depending on twist angle $\alpha$~\cite{2018arXiv181208097X},
\begin{align}\label{eq:tightbinding}
	H_0=-t(\alpha)\sum_{\langle ij\rangle}\sum_{\sigma=\uparrow,\downarrow}\sum_{o=1,2}c^\dagger_{i\sigma o}c_{j\sigma o}+\mathrm{h.c.}\,.
\end{align}
Here, the $c_{i\sigma o}^{(\dagger)}$ are the fermion annihilation (creation) operators on site $i$ with spin projection $\sigma$ and orbital (or band) index $o$. We add a chemical potential $\mu\sum_{i, \nu} n_{i \nu}$ with density $n_{i\nu}=c^\dagger_{i\nu}c_{i\nu}$.
Combining the indices $\sigma$ and $o$ as $\nu=(\sigma,o)$ into four flavors of fermion states makes an SU(4) symmetry of Eq.~\eqref{eq:tightbinding} apparent. 
In TBG and trilayer graphene/hBN, the index $o$ corresponds to the two valleys of the untwisted lattice system\cite{PhysRevX.8.031087,PhysRevX.8.031088,PhysRevX.8.031089}. But the model can also be considered in the more general context of multiorbital effects.
We expect spin-SU(2) symmetry to be a good approximation because of the small spin-orbit coupling of the light atoms composing graphene- or boron-nitrite-based systems. Approximate orbital degeneracy is assumed because in twisted graphene heterostructures, mixing of different valleys is suppressed by their large separation in momentum space~\cite{PhysRevX.8.031087}. In twisted hBN, \emph{ab initio} calculations even show an exact degeneracy\cite{2018arXiv181208097X}. 

To study correlations, we start with SU(4)-invariant Hubbard repulsion $U$ and nearest-neighbor exchange $J$
\begin{align}\label{eq:ia0}
	H_I=\frac{U}{2}\sum_{i,\nu,\nu'}n_{i\nu}n_{i\nu'} + J\sum_{\langle ij\rangle}\sum_{a=1}^{15}\hat{T}_i^a\hat{T}_j^a \,,
\end{align}
with $\hat T^a_i=c_{i \nu}^\dagger T^a_{\nu \nu'} c_{i\nu'}$. The $T^a_{\nu\nu'}$ are 4$\times$4 matrices forming the fundamental representation of the SU(4) Lie algebra.
Our motivation to consider the exchange $J$ is twofold. First, it was shown that nearest-neighbor interactions can be sizable despite the large moir\'e lattice spacing~\cite{PhysRevX.8.041041}. Second, the exchange term is introduced in the strong coupling limit of the onsite term. Being limited to weak and intermediate interactions by our method, we can use $J$ as a hint for the effect of stronger couplings.

We account for corrections to the approximate SU(4) symmetry by adding a Hund's coupling  $V_h$ and its equivalent in orbital space $K$, which can be induced from integrating out higher-energy phonons in an anti-adiabatic limit due to the small band widths of the flat bands \cite{PhysRevB.98.075154},
\begin{align}\label{eq:ia02}
	H_I'=-V_h\sum_i\vec{S}_i^2-K\sum_i\vec{L}_i^2\,,
\end{align}
where $\vec{S}_i=\frac{1}{2}c^\dagger_{i\sigma o}\vec{\sigma}_{\sigma\sigma'}c_{i\sigma' o}$ and  $\vec{L}_i=\frac{1}{2}c^\dagger_{i\sigma o}\vec{\tau}_{o o'}c_{i\sigma o'}$ with the Pauli matrices $\sigma_a, \tau_a, a \in \{1,2,3\}$. Summation over repeated spin and orbital indices is implied here and in the following~\footnote{The couplings in Eqs.~\eqref{eq:ia0}, \eqref{eq:ia02} relate to the common spin-SU(2) symmetric Hubbard and Hund terms $H_I=1/2\sum_{\sigma\sigma'}[\sum_{o}U_{\mathrm{SU(2)}} n_{io\sigma}n_{io\sigma'} + U'_{\mathrm{SU(2)}} \sum_{o\neq o'} n_{io\sigma}n_{io'\sigma'}+J_{\mathrm{SU(2)}}\sum_{o\neq o'}c^\dagger_{io\sigma}c^\dagger_{io'\sigma'}c_{io\sigma'}c_{io'\sigma'}]$ via $U_{\mathrm{SU(2)}}=U+V/2+K/2$, $U'_{\mathrm{SU(2)}}=U_{\mathrm{SU(2)}}$ and $J_{\mathrm{SU(2)}}=V-K$. Pair-hopping $J_{\mathrm{SU(2)}}'$ and $U_{\mathrm{SU(2)}}\neq U'_{\mathrm{SU(2)}}$ would break the SU(2)$\times$SU(2) symmetry. The term $\propto V$ can be absorbed by a shift into the terms $\propto U, K$~\cite{PhysRevB.98.075154}: $U\rightarrow U-V$, $K\rightarrow K+V$.}.
Alternatively, we consider nearest-neighbor spin or orbital SU(2) exchange couplings
$H_I''=\sum_{\langle ij\rangle}(J_s\vec{S}_i\vec{S}_j + K_n\vec{L}_i\vec{L}_j)\,$
to break the SU(4) symmetry and amplify the effect of nearest-neighbor exchange.
Upon inclusion of $H_I',H_I''$, a SU(2)$\times$SU(2) invariance remains, which originates from the SU(2) spin-rotational invariance and the two-orbital structure of the Hamiltonian.
Below, we assume that $V_h,K,J_s$ and $K_n$ are small compared to $U,J$, but we will comment on larger corrections to SU(4) in the 
discussion.

\begin{figure}[t!]
\includegraphics[width=0.85\columnwidth]{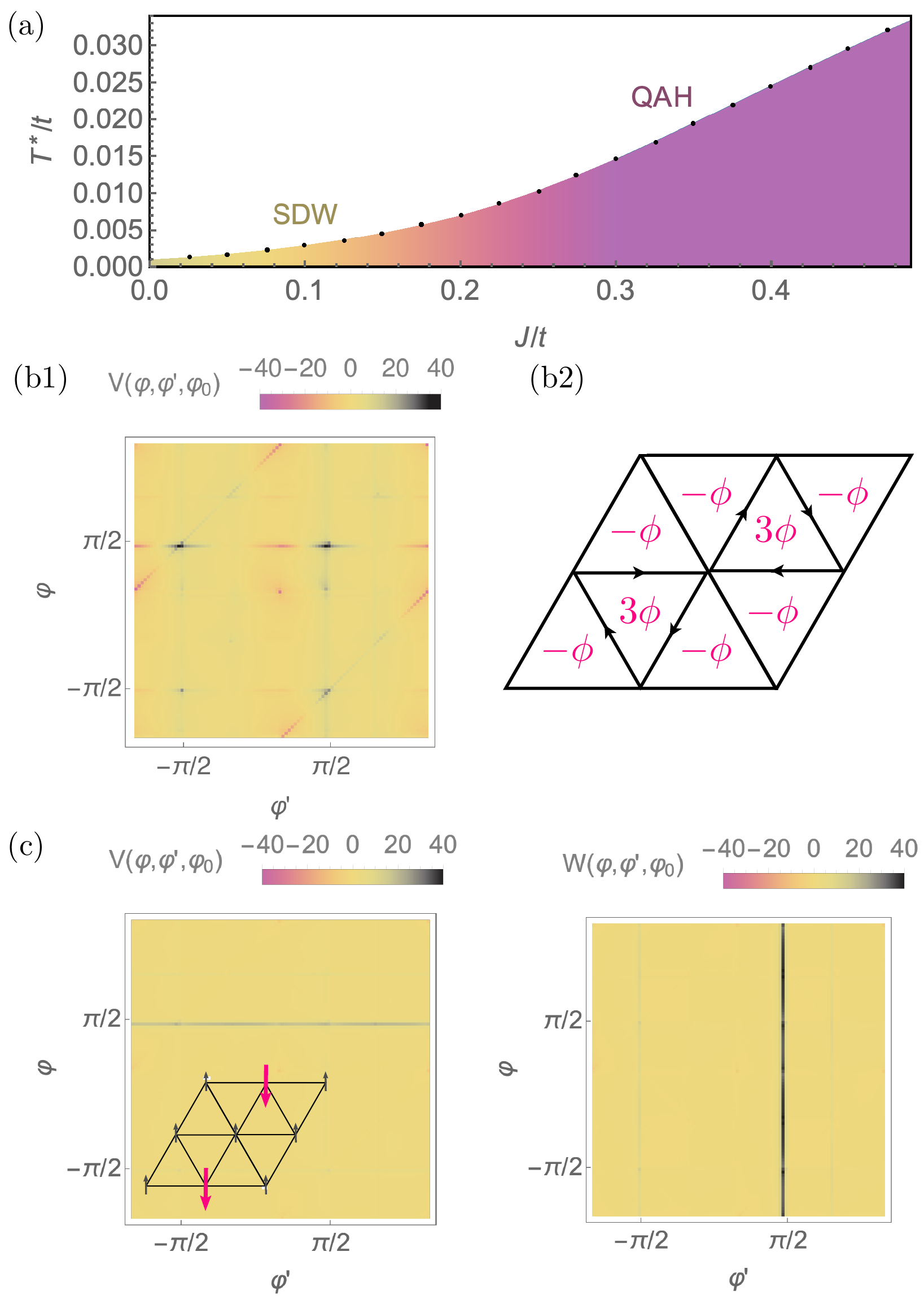}
\caption{(a) Instability temperature for initial $U=3t, V_h=0.4t, \mu=2t$ and variable SU(4) exchange $J$. For small $J$, there is a weak tendency towards a spin density wave (SDW), for increasing $J$ an interaction-induced quantum anomalous Hall state (QAH) develops (b1) The characteristic vertex $V(\varphi,\varphi',\varphi_0)$ at $T^*$ of the QAH instability. Initial couplings are $U=3t$, $J=0.4t$, $V_h=0.1t$ and $\mu=2t$. The angle of the third wavevector is fixed at $\varphi_0=-5\pi/6.$ The structure at $\varphi\approx \pi/2$ can be fitted by a d-wave form factors as given above Eq.~\eqref{eq:qah}. (b1) Sketch of the real space flux pattern of the QAH. (c) The characteristic vertices of the SDW instability for initial $U=3t$, $J=0$, $V_h=0.4t$ and $\mu=2t$. The inset shows the spin configuration of the triple-M SDW.
}
\label{fig:pd}
\end{figure}

\textit{Method.} 
%
We employ a functional renormalization group (fRG) approach for the one-particle-irreducible interaction vertices of the fermionic many-body system in the Fermi-surface patching scheme~\cite{RevModPhys.84.299,doi:10.1080/00018732.2013.862020}. Here, we specifically use that any SU(2)$\times$SU(2)-invariant interaction can be decomposed into two parts
\begin{align}
	\tilde H_I=& \sum_{k_1,k_2,k_3}\big[V(k_1,k_2,k_3)c^\dagger_{k_3\sigma o}c^\dagger_{k_4\sigma' o'}c_{k_2\sigma' o'}c_{k_1 \sigma o}\nonumber\\
	&+W(k_1,k_2,k_3)c^\dagger_{k_3\sigma o}c^\dagger_{k_4\sigma' o'}c_{k_2\sigma' o}c_{k_1 \sigma o'} \big]\,, \label{eq:ia}
\end{align}
where the first interaction vertex $\propto V$ is SU(4) invariant and the second one $\propto W$ breaks it down to SU(2)$\times$SU(2). The $k_i,\, i \in \{1,2,3,4\}$ are wavevectors in the first Brillouin zone (BZ) and $k_4$ is fixed by wavevector conservation. 
The fRG approach introduces an infrared cutoff $\Lambda$ and determines the renormalization of the system with respect to $\Lambda$. 
We derive fRG flow equations for the scale-dependent vertices $V$ and $W$ in the SU(2)$\times$SU(2)-invariant case, see App.~\ref{app:frg}. The fRG flow is initialized at the bandwidth of the system and integrated out toward the Fermi level upon lowering~$\Lambda$. This scheme amounts to an infinite-order summation of coupled particle-particle and particle-hole channels of second order in the effective interactions. It enables an unbiased investigation of the competing correlations through the analysis of the components of $V(k_1,k_2; k_3)$ and $W(k_1,k_2; k_3)$, which signal Fermi-liquid instabilities by flowing to large values at a critical fRG scale~$\Lambda_c$.
We use $\Lambda_c$ as an estimate for transition temperatures $\Lambda_c=T^*$. In a RPA-like summation, $T^*$ would equal the mean-field transition temperature, whereas it can be altered in our case due to the mutual feedback of the different channels. 
For the numerical computation, we discretize the wavevector dependence of $V$ and $W$. To this end, we project all wavevectors onto the Fermi surface and resolve the angular dependence by dividing the BZ into $N$ patches as shown in Fig.~\ref{fig:regions}, i.e. $V(k_1,k_2,k_3)\rightarrow V(\varphi_1,\varphi_2,\varphi_3)$ with $|k_i|=k_F(\varphi_i)$ being the corresponding Fermi vector.
In the following, we discuss our findings from varying $\mu$ and $J$, whilst keeping fixed $U, V_h$ and $K$ (or $J_s$ and $K_n$). Phase diagrams suggested by the diverging correlations are shown in Figs.~\ref{fig:regions}, \ref{fig:pd} and \ref{fig:fwave}. 

\textit{Instabilities of the nested Fermi surface. }
In our model, a chemical potential of $\mu=2t$ (three electrons per site) leads to a perfectly nested Fermi surface with three inequivalent nesting vectors $M_a$, $a\in\{1,2,3\}$ (see Fig.~\ref{fig:regions}). As a result, we find both real and imaginary density wave instabilities near van-Hove filling depending on the initial interactions. We assume that the onsite interaction $U$ is always dominating.

The SU(4) exchange coupling $J$ induces an ordering tendency in the imaginary charge density wave channel, which corresponds to a quantum anomalous Hall (QAH) state. It robustly appears in the tentative fRG phase diagram for varying initial couplings and persists when doping slightly away from van-Hove filling, $\mu\neq 2t$, cf. Fig.~\ref{fig:regions}. 
In particular, it is stable upon inclusion of small SU(4)-breaking terms.
The characteristic vertex structure of the QAH instability exhibits a special sign modulation, cf. Fig.~\ref{fig:pd}. 
The divergence only occurs in vertex $V$ for scattering processes connected by the different $M_a$ vectors. 
At van-Hove filling, the vertex can be fitted by $V(k_1,k_2,k_3)=\bar g\, d_a(k_1) d_a(k_2) \delta(k_3-k_1-M_a)$ with three $d$-wave functions for each of the $M$ points $d_1=2\sin(k_x/2)\sin(\sqrt{3}k_y/2), d_{2,3}=\mp \cos(k_x)\pm \cos((k_x\pm\sqrt{3}k_y)/2)$. 
This leads to the effective Hamiltonian
\begin{align}\label{eq:qah}
H_{\mathrm{QAH}}=\bar g\, \sum_{a} b^\dagger_{a}b_{a}\,,
\end{align}
where $b_a=\sum_k d_a(k) c_{k+M_a\sigma o}^\dagger c_{k\sigma o}$ and $\bar g>0$. Away from $\mu=2t$, the QAH divergencies are more limited to the vicinity of the $M$ points.
$H_{\mathrm{QAH}}$ suggests a purely imaginary mean-field configuration with $\langle b_{a}\rangle=-\langle b_{a}^\dagger\rangle$, leading to loop currents in real space~\cite{PhysRevB.93.115107,lin2019chiral}. We find simultaneous ordering tendencies of all three components $b_a$. This is expected to be energetically favored as it leads to a fully gapped state, which breaks time-reversal symmetry and has a non-zero Chern invariant\cite{PhysRevB.93.115107}. We show the corresponding flux configuration in Fig.~\ref{fig:pd}~(b).

For spin or orbital Hund couplings $V_h,K\neq0$ larger than the exchange $J$,  strong correlations of real SU(2)$\times$SU(2) spin/orbital-density waves (SODW) are induced. This is  signaled by sharp horizontal and vertical lines in the vertices as illustrated in Fig.~\ref{fig:pd} (c).
It depends on the SU(4)-breaking terms \eqref{eq:ia02}, which spin and orbital correlations are chosen from the allowed SU(2)$\times$SU(2) possibilities: $\vec{S}\cdot\vec{S}$, $\vec{L}\cdot\vec{L}$ or $(c^\dagger \vec{\sigma}\otimes\vec{\tau} c)\cdot (c^\dagger \vec{\sigma}\otimes\vec{\tau} c)$. For the example in Fig.~\ref{fig:pd} (b), we choose $U=3t, V_h=0.4t$ and $K=0$, which favors $\vec{S}\cdot\vec{S}$. We show other examples in the appendix.
The divergent lines in Fig.~\ref{fig:pd} again correspond to scattering processes that involve points on the Fermi surface connected by one of the three nesting vectors~$M_a$, cf. Fig.~\ref{fig:regions}. 
We read off the strength of the vertices as $V(k_1,k_2,k_3)=g\,\delta(k_3-k_1-M_a)$ and $W(k_1,k_2,k_3)=2g\,\delta(k_3-k_2-M_a)$ with $g>0$. 
This tensor structure  corresponds to the effective Hamiltonian
\begin{align}\label{eq:sodw}
H_{\mathrm{SDW}}^a&= g\, \vec{S}_{M_a} \cdot \vec{S}_{-M_a}\,,
\end{align}
with $\vec{S}_{M_a}=\sum_kc^\dagger_{k+M_a o \sigma }\vec{\sigma}_{\sigma \sigma'}  c_{k o\sigma' }$. 
The instability involves all three nesting vectors $M_a$ depending on which of them connects the respective part of the Fermi surface.
We interpret this as the itinerant triple-$M$ state of Ref.~\onlinecite{PhysRevLett.108.227204} for the triangular lattice which is conjectured to occur before an insulating chiral SDW sets in at lower temperatures~\cite{0295-5075-97-3-37001,PhysRevB.85.035414,PhysRevLett.101.156402}. We sketch the real-space spin pattern in the inset of Fig.~\ref{fig:pd} (c).
In contrast to the QAH, the SDW is relatively weak and quickly suppressed by tuning $\mu$ away from perfect nesting or increasing $J$. We attribute this to the increased available phase space due to the three inequivalent nesting vectors. Moreover, the higher flavor number favors the QAH, cf. Ref.~\onlinecite{lin2019chiral}. 
%

\textit{Unconventional superconductivity.}  
%
When we tune the chemical potential further away from perfect nesting, we find that the system is susceptible to superconductivity (SC) in the $d$-wave channel. In addition, we see strong pairing correlations at $\mu=2t$ for small onsite $U$ or large exchange $J$. Interestingly, the orbital degrees of freedom open the possibility to $d$-wave pairing with (spin-triplet)-(orbital-singlet) or (spin-singlet)-(orbital-triplet) symmetry. In our calculation it depends on the size of Hund's and orbital-Hund's couplings which of them is selected. 
We find that for $V_h>K$, superconductivity is mediated by fluctuations of the effective spin Hamiltonian Eq.~\eqref{eq:sodw}. 
Mean-field decoupling of Eq.~\eqref{eq:sodw} in the pairing channel with even parity leads to attraction in the (spin-singlet)-(orbital-triplet) channel. This is different to the strong-coupling scenario, where the Hund's coupling favors (spin-triplet)-(orbital-singlet) pairing. The reason is that in this case, superconductivity is mediated by purely ferromagnetic spin-fluctuations, which is qualitatively different from the mediation in terms of the weak-coupling spin-density wave as found here. Note, however, that one still can get (spin-triplet)-(orbital-singlet) $d$-wave SC from weak coupling, e.g., for $K>V_h>0$, an orbital density wave is induced for small $J$ mediating (spin-triplet)-(orbital-singlet) $d$-wave pairing.

\begin{figure}[t!]
\includegraphics[width=0.85\columnwidth]{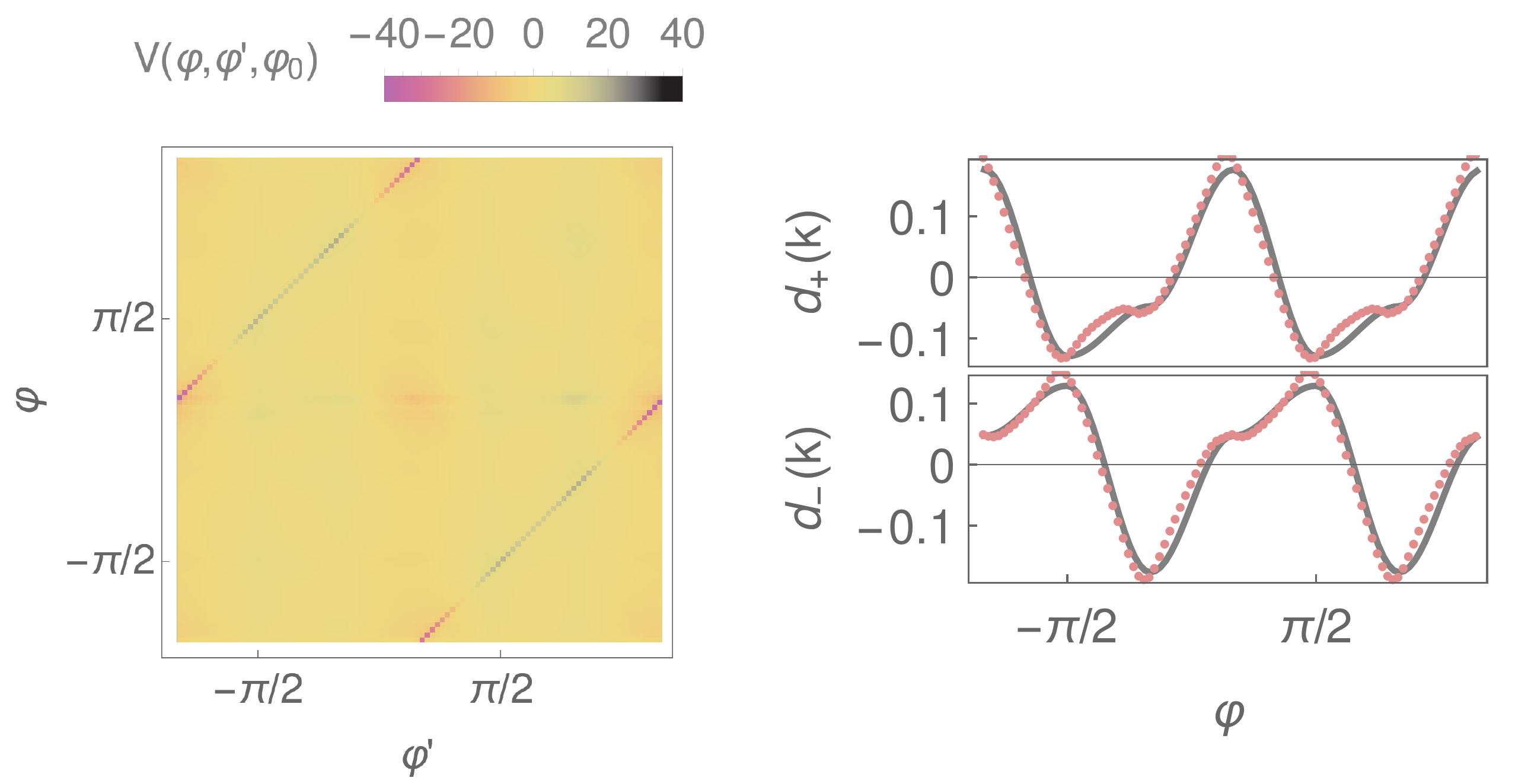}
\caption{Pairing vertex at the critical temperature for $\mu=1.9t$ and initial $U=3t$, $J=0.4t$, $V_h=0.1t$. The third angle $\varphi_0=-5\pi/6$. The extracted form factors on the right are fitted by a linear combination of $d_{x^2-y^2}$ and $d_{xy}$.}
\label{fig:SC}
\end{figure}

\begin{figure}[t!]
\includegraphics[width=0.85\columnwidth]{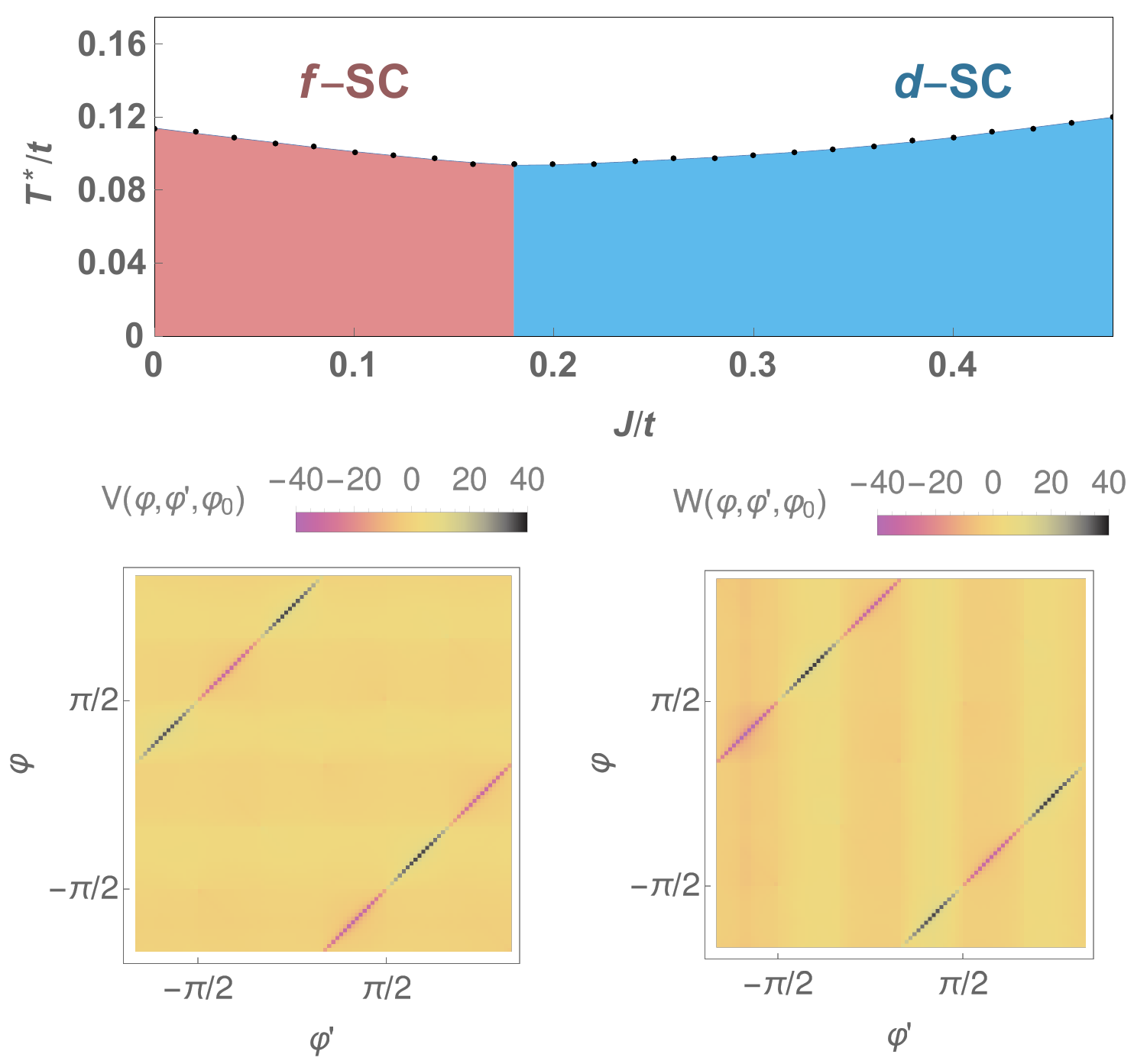}
\caption{(a) Instability temperature for $U=3t$, $V_h=0$ and  $J_s=1.2t$ with variable $J/t$. 
(b) Snapshot of the vertex for an instability towards $f$-wave superconductivity. Initial couplings are $U=3.0t, J_s=1.2t$ and $\mu=2.05t$. The angle of the third wavevector is set to $\phi_0= -2\pi/3$ Note that the diagonal structure crosses zero six times as opposed to four times for the $d$-wave instability in Fig.~\ref{fig:SC}.
}
\label{fig:fwave}
\end{figure}

A snapshot of the SC vertex is shown in Fig.~\ref{fig:SC} for the example of $\mu=1.9t$, $U=3t$, $J=0.4t$ and $V_h=0.1t$. It exhibits a dominant diagonal feature with $d$-wave form factor $d(k)$, i.e. $V(k_1,k_2,k_3)=-\hat g\, d(k_1) d(k_3) \delta(k_1+k_2)$ where $\hat g>0$. As we explain in App.~\ref{app:exham}, the relative sign between vertices $V$ and $W$ gives information about the spin and orbital pairing configuration.
In the above example, the divergent structure in $W(\varphi,\varphi',\varphi_0)$ (not shown) is weaker and has the same sign as compared to $V(\varphi,\varphi',\varphi_0)$. Therefore, the leading SC instability corresponds to (spin-singlet)-(orbital-triplet) pairing as expected. 
We extract the effective Hamiltonian
\begin{align}\label{eq:sc}
H_{\mathrm{SC}}=-\hat g\, \vec{\Delta}_d^\dagger \vec{\Delta}_d\,,
\end{align}
with $\vec{\Delta}_d=\sum_k d(k) c_{-k\sigma o} (i\vec{\tau}\tau^y)_{oo'} (i\sigma^y)_{\sigma\sigma'} c_{k\sigma'o'}$.
From our calculation, the form factor $d(k)$ is predicted to be a superposition of the two $d$-wave functions $d_{xy}=-2\sqrt{3}\sin k_x/2\cos \sqrt{2}k_y/2$ and $d_{x^2-y^2}=2\cos k_x-2\cos k_x/2\cos\sqrt{3}k_y/2$ (see Fig.~\ref{fig:SC}).
Eventually, we expect the $d_{x^2-y^2}\pm i d_{xy}$ states of Ref.~\onlinecite{PhysRevLett.121.087001} -- or rather their spin-singlet, orbital-triplet equivalent -- to be favored because this maximizes the pairing gap on the Fermi surface~\cite{0953-8984-26-42-423201} (see also Refs.~\onlinecite{PhysRevB.68.104510,nandkishore2012chiral}). As outlined in Ref.~\onlinecite{PhysRevLett.121.087001}, this leaves two types of $d\pm id$ superconductors degenerate at the mean-field level, which are both topological with protected edge states. Note that nematic superconductivity has also been discussed as an alternative\cite{PhysRevB.98.245103}.

Regarding the interplay with the QAH, we observe two different scenarios. In the case of a closed Fermi surface, $\mu<2t$, there appears to be a competition between both phases.
This is suggested by the behavior shown in Fig.~\ref{fig:regions} which exhibits a downturn of the superconducting instability temperature just before the steep increase very close to $\mu=2t$ when the QAH tendencies grow strong. 
In contrast, for $\mu\geq2t$, the diverging structure of the vertex suggest a crossover from a QAH- to SC-dominated regime when increasing $\mu$. In the intermediate regime, correlations in both channels grow large and one has to go beyond our approach to decide whether or not there will be a phase of coexistence.

We also find the possibility of $f$-wave SC with (spin-singlet)-(orbital-singlet) configuration. 
This pairing instability occurs for an open free Fermi surface, $\mu>2t$, and dominant $J_s>0$ or dominant $K_n>0$, see Fig.~\ref{fig:fwave}.  The unconventional combination of an odd form factor with spin-singlet is due to mediation by spin fluctuations: mean-field decoupling of Eq.~\eqref{eq:sodw} in the odd parity channels leads to attraction in the (spin-singlet)-(orbital-singlet) channel.
Our vertex data is well fitted by $f_{x(x^2-3y^2)}=\sin k_x-2\sin k_x/2\cos \sqrt{3}k_y/2$ (see Fig.~\ref{fig:fff}) with nodes along the $\Gamma-M_a$ directions. Thus, they do not coincide with the Fermi lines which are centered around the $K, K'$ points.
Since the gap function follows the form factor, this suggests a nodeless $f$-wave SC state.

\textit{Conclusion.} 
%
In this work, we have studied the quantum many-body instabilities of a multi-orbital model for interacting electrons on the triangular lattice. We have incorporated various universal aspects which were suggested to be relevant for the correlated behavior in  moir\'e heterostructures, such as twisted bilayer boron nitride and trilayer graphene-boron nitride systems.
We used the $N$-patch functional renormalization group approach for a  vertex structure with SU(2)$\times$SU(2) symmetry. This formulation of the fRG is a suitable method to investigate the interplay between (approximate) nesting and a large density of states in an unbiased way because it takes into account all correlation channels and their mutual couplings.
We have identified three classes of strongly-growing correlations: spin/orbital density wave states, two types of unconventional superconductivity and a Haldane-like loop-current phase with non-zero Chern number.

While the density waves are fragile and found only very close to perfect nesting, the superconducting and loop-current phases are robust, and occupy an extended range in parameter space. 
The type of pairing that emerges depends on the Fermi surface topology. 
We provided further evidence for the previously discussed topological $d\pm id$ superconductivity scenario~\cite{PhysRevLett.121.087001,lin2019chiral}. 
On the other hand, for an open Fermi surface the pairing correlations can be of $d$- or $f$-wave form in our computation, depending on the ratio and size of SU(4)-breaking Hund's or spin/orbital exchange couplings.
For larger corrections to the SU(4) invariance we find other types of instabilities. For example, slightly away from van Hove filling for $V_h>t$ and the other interactions set to zero, an (orbital-singlet)-(spin-triplet) SC instability with $s$-wave form factors emerges. For negative $J_s,K_n$, the flow is towards a ferromagnetic spin or ferro-orbital order. 

As a perspective for future studies we note that the fRG approach has the advantage that all modes of the effective model are taken into account. Hence, in principle, the resulting instability scales $T^*$ can serve as estimates for actual (short-range) ordering scales in the experimental system if the model parameters are known. At present, there are only rough estimates for the interaction parameters in TBG\cite{PhysRevX.8.031087} and twisted bilayer boron nitride\cite{2018arXiv181208097X} that point to potentially very large onsite repulsions $\sim 30\,\mathrm{meV}$ compared to effective hoppings $t\sim 0.33\,\mathrm{meV}$\cite{PhysRevX.8.031087}. The effect of these large onsite terms might, however, be reduced by the competition with nonlocal interactions~\cite{PhysRevLett.111.036601}, which are also expected to be sizeable. A study of this interplay is beyond the scope of our paper. Here we only state that the mentioned $t$-value, our critical scales for $U=3t$ end up in the sub-Kelvin range, $T^* \sim 0.03t \approx 0.1$K. Note, however, that these scales depend exponentially on the chosen couplings such that energy scales of a few Kelvin can be reached already for moderate interaction strengths.

As the discussion about the role of interactions is still ongoing, 
we think that the weak-to-intermediate coupling perspective given here, together with the use of the universal aspects of moir\'e heterostructures can provide essential insights into the general ordering tendencies in such multi-orbital systems.
In particular, our prediction of the appearance of robust topological/chiral interaction-induced phases is accessible to experimental examination.

\paragraph*{Acknowledgments.}
We thank Alexei Tsvelik for discussions.
M.M.S. was supported by the DFG through the Collaborative Research Center SFB1238, TP~C04.
L.C. acknowledges funding from the Alexander-von-Humboldt foundation. Work at BNL is supported by the U.S. Department of Energy, Office of Basic Energy Sciences, Contract No.~DE-SC0012704.
C.H. acknowledges funding
via the DFG research training group 1995 Quantum Many-Body Methods in Condensed Matter Systems.

\begin{appendix}

\section{Fermion functional RG setup}\label{app:frg}

We study many-body instabilities of the model defined in Eqs.~\eqref{eq:tightbinding}, \eqref{eq:ia0} and \eqref{eq:ia02} by means of the fRG method~\cite{Wetterich:1992yh}. To that end, we employ its specific formulation for correlated fermion systems~\cite{PhysRevB.63.035109,doi:10.1143/PTP.105.1}, see Refs.~\onlinecite{RevModPhys.84.299,doi:10.1080/00018732.2013.862020} for  reviews.
The fRG-$N$-patch scheme starts with a fermionic action corresponding to our model Hamiltonian
$
S[\bar\psi,\psi] = -(\bar \psi,G_0^{-1} \psi) + \mathcal{V}[\bar \psi,\psi]\,.
$
The quadratic first term with the free propagator
$
G_0(\omega_n,k)= 1/(i \omega_n - \epsilon(k))
$
includes Matsubara frequencies $\omega_n$ and wavevectors $k$.
The dispersion $\epsilon(k)$ is measured relative to the chemical potential.
In the present case, the free propagator is diagonal with respect to spin and the orbital quantum numbers, $\sigma$ and $o$, and we have suppressed the according indices.
The general interaction contribution $\mathcal{V}[\bar\psi,\psi]$ is quartic in the fermion fields. Its specific form can be directly inferred from the Hamiltonians in Eqs.~\eqref{eq:ia0} and~\eqref{eq:ia02}.

The bare propagator is then regularized by an infrared cutoff with energy scale~$\Lambda$:
$
G_0(\omega_n,k) \to G_0^{\Lambda}(\omega_n,k) = \theta_{\varepsilon}^{\Lambda}(\epsilon(k))/(i \omega_n - \epsilon(k))\,,
$
where $\theta_{\varepsilon}^{\Lambda}$ is an approximate step function with smoothening scale $\varepsilon$ cutting off fluctuations with energies $|\epsilon(k)| \lesssim \Lambda$. 
The regularized propagator $G_0^{\Lambda}$ is then used to set up the functional integral for the scale-dependent effective action $\Gamma^\Lambda$, which generates the one-particle irreducible vertex functions $\Gamma^{(2n)\Lambda}$.

The fRG flow is generated upon variation of $\Lambda$ producing a hierarchy of differential equations for the vertex functions $\Gamma^{(2n)\Lambda}$. Integration of the flow towards the infrared $\Lambda\rightarrow 0$ yields the full effective action $\Gamma$.
We employ the standard truncation for analyzing many-body instabilities in two-dimensional fermion systems, where the fRG flow of all higher $n$-point functions with $n \geq 6$ and self-energy feedback are neglected.
In our SU(2)$\times$SU(2)-invariant system, this corresponds to following the scale dependence of the two generally frequency- and momentum-dependent effective interaction vertices $V$ and $W$ as defined in Eq.~\eqref{eq:ia}. 
Their flow consists of three contributions, i.e. the particle-particle, direct and crossed particle-hole channels:
\begin{widetext}
\vspace{-0.5cm}
\begin{align}
\partial_\Lambda V(q_1,q_2,q_3)&= T_{pp}^V(q_1,q_2,q_3) +T_{ph,d}^V(q_1,q_2,q_3) + T_{ph,cr}^V(q_1,q_2,q_3)\label{eq:flow01}\\
\partial_\Lambda W(q_1,q_2,q_3)&= T_{pp}^W(q_1,q_2,q_3) +T_{ph,d}^W(q_1,q_2,q_3) + T_{ph,cr}^W(q_1,q_2,q_3) \,.
\end{align}
The expression for the different channels are
\begin{align}
T_{pp}^V(q_1,q_2,q_3)&=-\int dk L_-(q_1+q_2,k)\Big[ V(q_{pp},k,q_3) V(q_1,q_2,q_{pp}) + W(q_{pp},k,q_3) W(q_1,q_2,q_{pp})\Big] \\
T_{pp}^W(q_1,q_2,q_3)&=-\int dk L_-(q_1+q_2,k)\Big[ V(q_{pp},k,q_3) W(q_1,q_2,q_{pp}) + W(q_{pp},k,q_3) V(q_1,q_2,q_{pp})\Big]\\
T_{ph,d}^V(q_1,q_2,q_3)&=-\int dk L_+(q_1-q_3,k)\Big[-4 V(q_1,k,q_3) V(q_{ph,d},q_2,k) +  V(k,q_1,q_3) V(q_{ph,d},q_2,k) \notag \\ &+ V(q_1,k,q_3) V(q_2,q_{h,d},k) + W(k,q_1,q_3) W(q_{ph,d},q_2,k) + W(q_1,k,q_3) W(q_2,q_{ph,d},k) \notag \\ &- 2 V(q_1,k,q_3) W(q_{ph,d},q_2,k)  + 2 V(q_1,k,q_3) W(q_2,q_{ph,d},k) - 2 W(q_1,k,q_3) V(q_{ph,d},q_2,k)  \notag \\ &+ 2 W(k,q_1,q_3) V(q_{ph,d},q_2,k) \Big]\\
T_{ph,d}^W(q_1,q_2,q_3)&=-\int dk L_+(q_1-q_3,k)\Big[-2 W(q_1,k,q_3) W(q_{ph,d},q_2,k) +  V(k,q_1,q_3) W(q_{ph,d},q_2,k) \notag \\ &+ W(q_1,k,q_3) V(q_2,q_{ph,d},k) \Big]\\
T_{ph,cr}^V(q_1,q_2,q_3)&=-\int dk L_+(q_2-q_3,k) V(k,q_2,q_3) V(q_1,q_{ph,cr},k) \\
T_{ph,cr}^W(q_1,q_2,q_3)&=-\int dk L_+(q_2-q_3,k)\Big[ 2W(k,q_2,q_3) W(q_1,q_{ph,cr},k) + V(k,q_2,q_3) W(q_1,q_{ph,cr},k) \notag \\ &+ W(k,q_2,q_3) V(q_1,q_{ph,cr},k) \Big]\label{eq:flow09} \,,
\end{align}
\end{widetext}
where we defined $q_{pp}=q_1+q_2-k$, $q_{ph,d}=q_1-q_3+k$, $q_{ph,cr}=q_2-q_3+k$ and $L_{\pm}(q,k)=S^\Lambda(k)G_0^\Lambda(q\pm k)+G_0^\Lambda(k)S^\Lambda(q\pm k)$ with the single-scale propagator $S^\Lambda=\partial_\Lambda G_0^\Lambda$.
In Eqs.~\eqref{eq:flow01}-\eqref{eq:flow09}, the arguments are combined frequency and momentum vectors $q=(\omega,\vec{q})$. 
The most singular part of the vertices is expected to come from zero Matsubara frequency. Thus, for the numerical evaluation of instabilities, we do not resolve the frequency dependence of the two-particle vertices and only consider the zero-frequency limit. The wavevector dependence is taken into account via a patching of the Fermi surface which resolves the angular direction. This truncation has been shown to successfully describe the Fermi surface instabilities of numerous systems. In particular, it goes beyond the random phase approximation because the coupling of the different channels is taken into account.
We discretize the Fermi surface by patch points, each representing one of $N$ patches covering the BZ, as shown in Fig.~\ref{fig:regions}. We use up to $N=96$ to check the convergence of our results, which is particularly important at van-Hove filling. For $\mu>2t$, the Fermi surface is open and we choose patches that are centered around the $K, K'$ points. For details on the patching, we refer to Ref.~\onlinecite{PhysRevB.98.045142} which depicts both geometries used here.

We initialize our numerics with an RG scale equal to the bandwidth. An instability towards an ordered state is signaled by a divergence in the two-particle vertices $V, W$ during the flow towards the infrared and we  stop the RG evolution when this occurs. In practice, our stopping condition is that one vertex component exceeds $40t$. When $V, W$ remain finite, we stop the flow when $\Lambda < 10^{-9}t$.
%

\section{Extraction of effective Hamiltonians}\label{app:exham}

The snapshots of the vertices in Figs.~\ref{fig:pd}~(b)-(d) show that the fRG predicts specific momentum dependencies for the many-body instabilities, which allows us to characterize the occuring ordering tendencies in some detail.
Therefore, we systematically extract~\cite{PhysRevB.90.045135} the effective Hamiltonians, cf. Eqs.~\eqref{eq:qah}, \eqref{eq:sodw}, \eqref{eq:sc}.

\begin{figure}[h!]
\includegraphics[width=0.9\columnwidth]{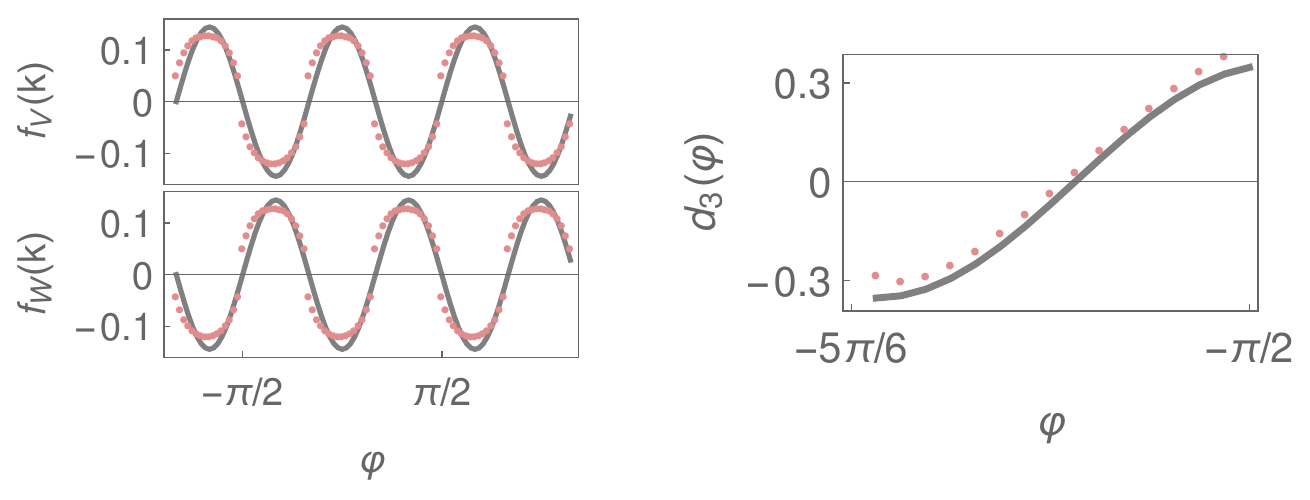}
\caption{Left: SC $f$-wave form factor of the $V$ and $W$ vertices shown in Fig.~\ref{fig:fwave} $f_V=-f_W=-\sin k_x+2\sin k_x/2\cos\sqrt{2}k_y/2$. Initial values are $U=3.0t, J_s=1.2t$ and $\mu=2.05t$. Right: d-wave form factor of the QAH instability of the vertex in Fig.~\ref{fig:pd}. We show $d_3=\cos (k_x)-\cos((k_x-\sqrt{3}k_y)/2)$ as example. }
\label{fig:fff}
\end{figure}

In particular, for the SC instability, the procedure is as follows. The diverging part of the vertex suggests $V(k_1,k_2,k_3)\rightarrow V(k,-k,k')$. We compute the eigensystem of this effective $N\times N$ matrix as defined on the patch points.
The eigenvector with the largest absolute value is expected to provide the order parameter with the highest transition temperature in a meanfield approach and we therefore examine its momentum dependence in more detail. 
In the case of the $d$-wave SC instability, we find two eigenvectors with degenerate largest eigenvalue.
Introducing the momentum-space representations of the $d$-wave form factors on the triangular lattice,
$d_{x^2-y^2}=2 \cos(k_x) - 2 \cos(k_x/2) \cos(\sqrt{3} k_y/2)$ and  
$d_{xy}=-2\sqrt{3} \sin(k_x/2) \sin(\sqrt{3}k_y)/2)$,
we can fit the extracted momentum dependencies very well by $d_+(k)=-\sin(\phi)d_{x^2-y^2}+ \cos(\phi)d_{xy}$ and $d_-(k)=\cos(\phi)d_{x^2-y^2}+ \sin(\phi)d_{xy}$ with $\phi=5\pi/4$ as shown in the Fig.~\ref{fig:SC}. 
We also show the $f$-wave form factor extracted from the vertices in Fig~\ref{fig:fwave} in Fig.~\ref{fig:fff}.

%

We now only keep these leading channels, i.e. $V(k,-k,k')\rightarrow \lambda^V d(k)^*d(k')$ with corresponding form factor superposition $d(k)$.
To connect to the standard notation, we use the Fierz identity $\delta_{ad}\delta_{bc}=(\boldsymbol{\sigma}_{ab}\boldsymbol{\sigma}_{cd}+\delta_{ab}\delta_{cd})/2=[(\Gamma_0^\dagger)_{ab}(\Gamma_0)_{cd}+ (\boldsymbol{\Gamma}^\dagger)_{ab}(\boldsymbol{\Gamma}^\dagger)_{cd}]/2$ with $\Gamma_0=i\sigma_2$ and $\Gamma_i=\sigma_i i \sigma_2$.
Accordingly, we define the matrices $G_0=i\tau_2$ and  $\vec{G}=\vec{\tau}i\tau_2$ in orbital space.
We rewrite the interaction
\begin{widetext}
\vspace{-0.5cm}
\begin{align}
V^{\mathrm{SC}}&=\frac{\lambda^V}{4} d^V(k)^*d^V(k') c_{ka\alpha}^\dagger c_{-kb\beta}^\dagger c_{-k'c\gamma} c_{k'd\delta}\left[ (G_0^\dagger)_{ab}(G_0)_{cd}+ (\boldsymbol{G}^\dagger)_{ab}(\boldsymbol{G})_{cd} \right] \otimes \left[ (\Gamma_0^\dagger)_{\alpha\beta}(\Gamma_0)_{\gamma\delta}+ (\boldsymbol{\Gamma}^\dagger)_{\alpha\beta}(\boldsymbol{\Gamma})_{\gamma\delta} \right] \notag \\
&\hspace{-0.5cm}+\frac{\lambda^W}{4} d^W(k)^*d^W(k') c_{ka\alpha}^\dagger c_{-kb\beta}^\dagger c_{-k'c\gamma} c_{k'd\delta}\left[ -(G_0^\dagger)_{ab}(G_0)_{cd}+ (\boldsymbol{G}^\dagger)_{ab}(\boldsymbol{G})_{cd} \right] \otimes\left[ (\Gamma_0^\dagger)_{\alpha\beta}(\Gamma_0)_{\gamma\delta}+ (\boldsymbol{\Gamma}^\dagger)_{\alpha\beta}(\boldsymbol{\Gamma})_{\gamma\delta} \right].
\end{align}
\end{widetext}
We used Greek symbols for  spin and Latin symbols for orbital space. The minus sign in the expression from the $W$ vertex is due to a transpose that one has to take because of the exchanged orbital index. Now, we see if $d^i(k),\ i \in \{V,W\}$ is even, only $G_0^\dagger G_0\otimes \boldsymbol{\Gamma}^\dagger \boldsymbol{\Gamma}$ or $\boldsymbol{G}^\dagger \boldsymbol{G}\otimes \Gamma_0^\dagger \Gamma_0$ are allowed by the antisymmetry of the pairing function. This correspond to (orbital-singlet)-(spin-triplet) and (orbital-triplet)-(spin-singlet) SC, respectively. Due to the minus sign in the expression, we find for $\lambda^V=\lambda^W$ (orbital-triplet)-(spin-singlet) and for $\lambda^V=-\lambda^W$ (orbital-singlet)-(spin-triplet).

For the QAH instability we proceed accordingly and find that the form factor $d_a$ is determined by the momentum transfer $M_a$ as given in the main text. We plot $d_3(k)$ as an example, see top panel in Fig.~\ref{fig:fff}.

\section{Other instabilities}
\begin{figure}[t!]
\includegraphics[width=0.85\columnwidth]{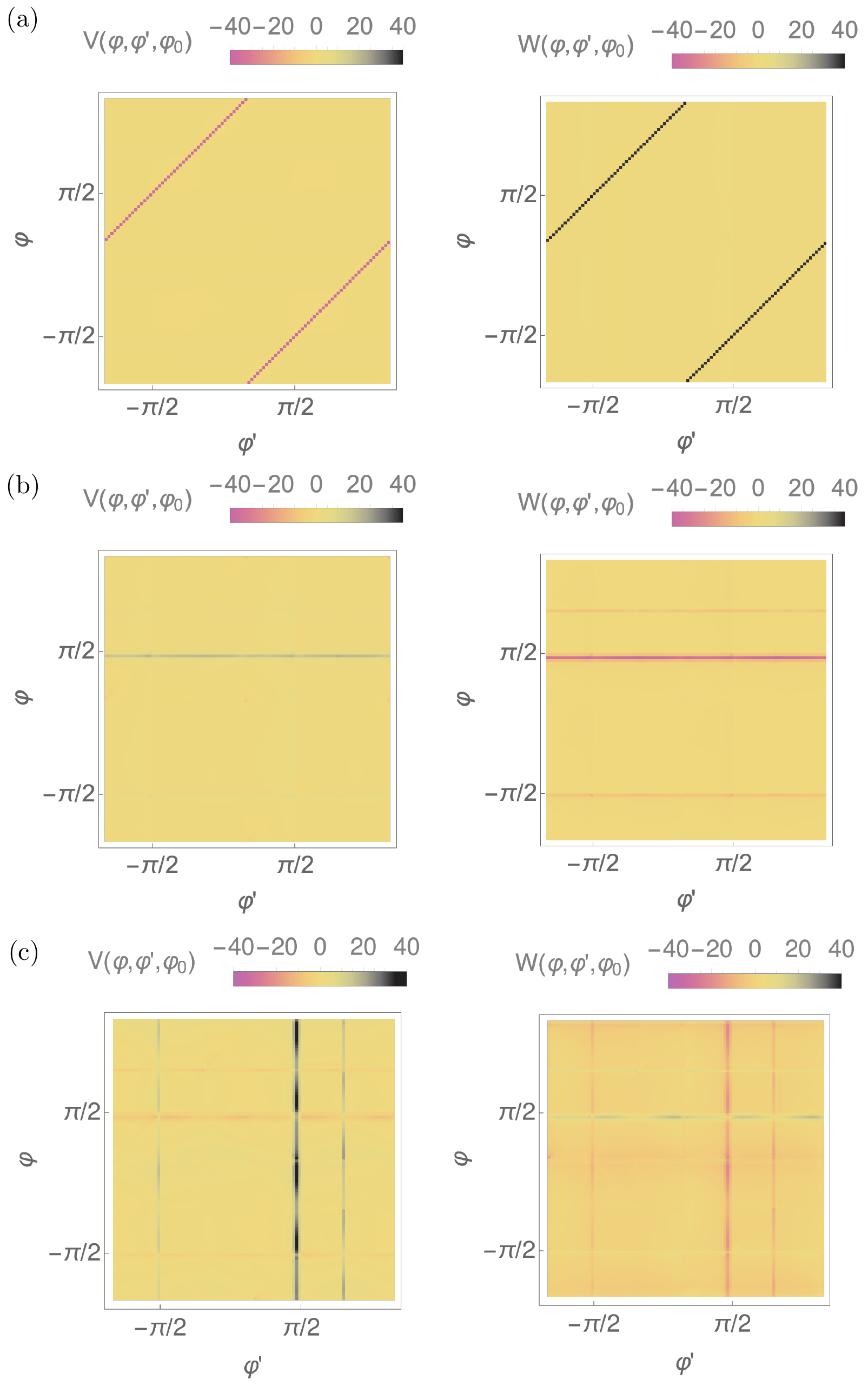}
\caption{(a) Pairing vertex with s-wave form factor for $V_h=1.2t$ and  $\mu=1.98t$.
(b) Vertex of an orbital density wave for $\mu=2t$ and initial $U=3t$ and $K=0.4t$
(c) Vertex of a spin-orbital density wave for $\mu=2t$, $U=3t$, $J_s=-0.3t$, $K_n=-0.4t$, $V_h=0.1t$. Non-specified couplings are zero.}
\label{fig:sSC}
\end{figure}

As we mentioned in the main text, we find an orbital-singlet, spin-triplet $s$-wave instability for small onsite repulsion and dominant Hund's coupling. The reason is that there is an attractive onsite interaction for this combination. We show the corresponding vertices in Fig.~\ref{fig:sSC}~(a).

We also note that besides the spin-density wave discussed in the main text, the system can flow towards an orbital density wave or a combined spin-orbital density wave at $\mu=2t$. It depends on the Hund-like $V,K$ and SU(2) exchange $J_s,K_n$ couplings which density-wave instability of the allowed SU(2)$\times$SU(2) combinations is chosen. We show examples for the vertex of an orbital density wave in Fig.~\ref{fig:sSC} (b) with initial couplings $U=3t$, $K=0.4t$. The leading vertex structure corresponds to 
\begin{align}\label{eq:osdw}
H_{\mathrm{ODW}}^a= g_{\mathrm{ODW}} \vec{L}_{M_a} \cdot \vec{L}_{-M_a}
\end{align}
with $\vec{L}_{M_a}=\sum_kc^\dagger_{k+M_a  o \sigma} \vec{\tau}_{oo'} c_{k  o'\sigma}$ and $g_{\mathrm{ODW}}>0$.
For $U=3t$ and $V=K=0$, but $J_s,K_n\neq0$, the density-wave instability corresponds to a combined spin-orbital density wave. That is, we find $V(k_1,k_2,k_3)=g_{\mathrm{SODW}}[4\delta(k_3-k_2-M_a)-\delta(k_3-k_1-M_a)]$ and $W(k_1,k_2,k_3)=2g_{\mathrm{SODW}}[-\delta(k_3-k_2-M_a)+\delta(k_3-k_1-M_a)]$ with $g_{\mathrm{SODW}}>0$ and the effective Hamiltonian becomes
\begin{align}\label{eq:osdw}
H_{\mathrm{OSDW}}^a= g_{\mathrm{SODW}} \vec{J}_{M_a} \cdot \vec{J}_{-M_a}
\end{align}
with $\vec{J}_{M_a}=\sum_kc^\dagger_{k+M_a \sigma o}\vec{\sigma}_{\sigma\sigma'} \vec{\tau}_{oo'} c_{k \sigma' o'}$.

\end{appendix}

\bibliography{Bib_moire}

\begin{thebibliography}{49}%
\makeatletter
\providecommand \@ifxundefined [1]{%
 \@ifx{#1\undefined}
}%
\providecommand \@ifnum [1]{%
 \ifnum #1\expandafter \@firstoftwo
 \else \expandafter \@secondoftwo
 \fi
}%
\providecommand \@ifx [1]{%
 \ifx #1\expandafter \@firstoftwo
 \else \expandafter \@secondoftwo
 \fi
}%
\providecommand \natexlab [1]{#1}%
\providecommand \enquote  [1]{``#1''}%
\providecommand \bibnamefont  [1]{#1}%
\providecommand \bibfnamefont [1]{#1}%
\providecommand \citenamefont [1]{#1}%
\providecommand \href@noop [0]{\@secondoftwo}%
\providecommand \href [0]{\begingroup \@sanitize@url \@href}%
\providecommand \@href[1]{\@@startlink{#1}\@@href}%
\providecommand \@@href[1]{\endgroup#1\@@endlink}%
\providecommand \@sanitize@url [0]{\catcode `\\12\catcode `\$12\catcode
  `\&12\catcode `\#12\catcode `\^12\catcode `\_12\catcode `\%12\relax}%
\providecommand \@@startlink[1]{}%
\providecommand \@@endlink[0]{}%
\providecommand \url  [0]{\begingroup\@sanitize@url \@url }%
\providecommand \@url [1]{\endgroup\@href {#1}{\urlprefix }}%
\providecommand \urlprefix  [0]{URL }%
\providecommand \Eprint [0]{\href }%
\providecommand \doibase [0]{http://dx.doi.org/}%
\providecommand \selectlanguage [0]{\@gobble}%
\providecommand \bibinfo  [0]{\@secondoftwo}%
\providecommand \bibfield  [0]{\@secondoftwo}%
\providecommand \translation [1]{[#1]}%
\providecommand \BibitemOpen [0]{}%
\providecommand \bibitemStop [0]{}%
\providecommand \bibitemNoStop [0]{.\EOS\space}%
\providecommand \EOS [0]{\spacefactor3000\relax}%
\providecommand \BibitemShut  [1]{\csname bibitem#1\endcsname}%
\let\auto@bib@innerbib\@empty
\bibitem [{\citenamefont {Cao}\ \emph {et~al.}(2018{\natexlab{a}})\citenamefont
  {Cao}, \citenamefont {Fatemi}, \citenamefont {Demir}, \citenamefont {Fang},
  \citenamefont {Tomarken}, \citenamefont {Luo}, \citenamefont
  {Sanchez-Yamagishi}, \citenamefont {Watanabe}, \citenamefont {Taniguchi},
  \citenamefont {Kaxiras} \emph {et~al.}}]{cao2018correlated}%
  \BibitemOpen
  \bibfield  {author} {\bibinfo {author} {\bibfnamefont {Y.}~\bibnamefont
  {Cao}}, \bibinfo {author} {\bibfnamefont {V.}~\bibnamefont {Fatemi}},
  \bibinfo {author} {\bibfnamefont {A.}~\bibnamefont {Demir}}, \bibinfo
  {author} {\bibfnamefont {S.}~\bibnamefont {Fang}}, \bibinfo {author}
  {\bibfnamefont {S.~L.}\ \bibnamefont {Tomarken}}, \bibinfo {author}
  {\bibfnamefont {J.~Y.}\ \bibnamefont {Luo}}, \bibinfo {author} {\bibfnamefont
  {J.~D.}\ \bibnamefont {Sanchez-Yamagishi}}, \bibinfo {author} {\bibfnamefont
  {K.}~\bibnamefont {Watanabe}}, \bibinfo {author} {\bibfnamefont
  {T.}~\bibnamefont {Taniguchi}}, \bibinfo {author} {\bibfnamefont
  {E.}~\bibnamefont {Kaxiras}},  \emph {et~al.},\ }\bibfield  {title}
  {{\color{Gray}\small \bibinfo {title} {Correlated insulator behaviour at
  half-filling in magic-angle graphene superlattices},\ }}\href@noop {}
  {\bibfield  {journal} {\bibinfo  {journal} {Nature}\ }\textbf {\bibinfo
  {volume} {556}},\ \bibinfo {pages} {80} (\bibinfo {year}
  {2018}{\natexlab{a}})}\BibitemShut {NoStop}%
\bibitem [{\citenamefont {Cao}\ \emph {et~al.}(2018{\natexlab{b}})\citenamefont
  {Cao}, \citenamefont {Fatemi}, \citenamefont {Fang}, \citenamefont
  {Watanabe}, \citenamefont {Taniguchi}, \citenamefont {Kaxiras},\ and\
  \citenamefont {Jarillo-Herrero}}]{cao2018unconventional}%
  \BibitemOpen
  \bibfield  {author} {\bibinfo {author} {\bibfnamefont {Y.}~\bibnamefont
  {Cao}}, \bibinfo {author} {\bibfnamefont {V.}~\bibnamefont {Fatemi}},
  \bibinfo {author} {\bibfnamefont {S.}~\bibnamefont {Fang}}, \bibinfo {author}
  {\bibfnamefont {K.}~\bibnamefont {Watanabe}}, \bibinfo {author}
  {\bibfnamefont {T.}~\bibnamefont {Taniguchi}}, \bibinfo {author}
  {\bibfnamefont {E.}~\bibnamefont {Kaxiras}}, \ and\ \bibinfo {author}
  {\bibfnamefont {P.}~\bibnamefont {Jarillo-Herrero}},\ }\bibfield  {title}
  {{\color{Gray}\small \bibinfo {title} {Unconventional superconductivity in
  magic-angle graphene superlattices},\ }}\href@noop {} {\bibfield  {journal}
  {\bibinfo  {journal} {Nature}\ }\textbf {\bibinfo {volume} {556}},\ \bibinfo
  {pages} {43} (\bibinfo {year} {2018}{\natexlab{b}})}\BibitemShut {NoStop}%
\bibitem [{\citenamefont {Yankowitz}\ \emph {et~al.}(2018)\citenamefont
  {Yankowitz}, \citenamefont {Chen}, \citenamefont {Polshyn}, \citenamefont
  {Watanabe}, \citenamefont {Taniguchi}, \citenamefont {Graf}, \citenamefont
  {Young},\ and\ \citenamefont {Dean}}]{yankowitz2018tuning}%
  \BibitemOpen
  \bibfield  {author} {\bibinfo {author} {\bibfnamefont {M.}~\bibnamefont
  {Yankowitz}}, \bibinfo {author} {\bibfnamefont {S.}~\bibnamefont {Chen}},
  \bibinfo {author} {\bibfnamefont {H.}~\bibnamefont {Polshyn}}, \bibinfo
  {author} {\bibfnamefont {K.}~\bibnamefont {Watanabe}}, \bibinfo {author}
  {\bibfnamefont {T.}~\bibnamefont {Taniguchi}}, \bibinfo {author}
  {\bibfnamefont {D.}~\bibnamefont {Graf}}, \bibinfo {author} {\bibfnamefont
  {A.~F.}\ \bibnamefont {Young}}, \ and\ \bibinfo {author} {\bibfnamefont
  {C.~R.}\ \bibnamefont {Dean}},\ }\bibfield  {title} {{\color{Gray}\small
  \bibinfo {title} {Tuning superconductivity in twisted bilayer graphene},\
  }}\href@noop {} {\bibfield  {journal} {\bibinfo  {journal} {arXiv preprint
  arXiv:1808.07865}\ } (\bibinfo {year} {2018})}\BibitemShut {NoStop}%
\bibitem [{\citenamefont {Chen}\ \emph {et~al.}(2018)\citenamefont {Chen},
  \citenamefont {Jiang}, \citenamefont {Wu}, \citenamefont {Lv}, \citenamefont
  {Li}, \citenamefont {Watanabe}, \citenamefont {Taniguchi}, \citenamefont
  {Shi}, \citenamefont {Zhang},\ and\ \citenamefont {Wang}}]{chen2018gate}%
  \BibitemOpen
  \bibfield  {author} {\bibinfo {author} {\bibfnamefont {G.}~\bibnamefont
  {Chen}}, \bibinfo {author} {\bibfnamefont {L.}~\bibnamefont {Jiang}},
  \bibinfo {author} {\bibfnamefont {S.}~\bibnamefont {Wu}}, \bibinfo {author}
  {\bibfnamefont {B.}~\bibnamefont {Lv}}, \bibinfo {author} {\bibfnamefont
  {H.}~\bibnamefont {Li}}, \bibinfo {author} {\bibfnamefont {K.}~\bibnamefont
  {Watanabe}}, \bibinfo {author} {\bibfnamefont {T.}~\bibnamefont {Taniguchi}},
  \bibinfo {author} {\bibfnamefont {Z.}~\bibnamefont {Shi}}, \bibinfo {author}
  {\bibfnamefont {Y.}~\bibnamefont {Zhang}}, \ and\ \bibinfo {author}
  {\bibfnamefont {F.}~\bibnamefont {Wang}},\ }\bibfield  {title}
  {{\color{Gray}\small \bibinfo {title} {{Gate-Tunable Mott Insulator in
  Trilayer Graphene-Boron Nitride Moir\'e Superlattice}},\ }}\href@noop {}
  {\bibfield  {journal} {\bibinfo  {journal} {arXiv preprint arXiv:1803.01985}\
  } (\bibinfo {year} {2018})}\BibitemShut {NoStop}%
\bibitem [{\citenamefont {Chen}\ \emph {et~al.}(2019)\citenamefont {Chen},
  \citenamefont {Sharpe}, \citenamefont {Gallagher}, \citenamefont {Rosen},
  \citenamefont {Fox}, \citenamefont {Jiang}, \citenamefont {Lyu},
  \citenamefont {Li}, \citenamefont {Watanabe}, \citenamefont {Taniguchi} \emph
  {et~al.}}]{chen2019signatures}%
  \BibitemOpen
  \bibfield  {author} {\bibinfo {author} {\bibfnamefont {G.}~\bibnamefont
  {Chen}}, \bibinfo {author} {\bibfnamefont {A.~L.}\ \bibnamefont {Sharpe}},
  \bibinfo {author} {\bibfnamefont {P.}~\bibnamefont {Gallagher}}, \bibinfo
  {author} {\bibfnamefont {I.~T.}\ \bibnamefont {Rosen}}, \bibinfo {author}
  {\bibfnamefont {E.}~\bibnamefont {Fox}}, \bibinfo {author} {\bibfnamefont
  {L.}~\bibnamefont {Jiang}}, \bibinfo {author} {\bibfnamefont
  {B.}~\bibnamefont {Lyu}}, \bibinfo {author} {\bibfnamefont {H.}~\bibnamefont
  {Li}}, \bibinfo {author} {\bibfnamefont {K.}~\bibnamefont {Watanabe}},
  \bibinfo {author} {\bibfnamefont {T.}~\bibnamefont {Taniguchi}},  \emph
  {et~al.},\ }\bibfield  {title} {{\color{Gray}\small \bibinfo {title}
  {{Signatures of Gate-Tunable Superconductivity in Trilayer Graphene/Boron
  Nitride Moir\'e Superlattice}},\ }}\href@noop {} {\bibfield  {journal}
  {\bibinfo  {journal} {arXiv preprint arXiv:1901.04621}\ } (\bibinfo {year}
  {2019})}\BibitemShut {NoStop}%
\bibitem [{\citenamefont {Fu}\ \emph {et~al.}(2018)\citenamefont {Fu},
  \citenamefont {K{\"o}nig}, \citenamefont {Wilson}, \citenamefont {Chou},\
  and\ \citenamefont {Pixley}}]{fu2018magic}%
  \BibitemOpen
  \bibfield  {author} {\bibinfo {author} {\bibfnamefont {Y.}~\bibnamefont
  {Fu}}, \bibinfo {author} {\bibfnamefont {E.}~\bibnamefont {K{\"o}nig}},
  \bibinfo {author} {\bibfnamefont {J.}~\bibnamefont {Wilson}}, \bibinfo
  {author} {\bibfnamefont {Y.-Z.}\ \bibnamefont {Chou}}, \ and\ \bibinfo
  {author} {\bibfnamefont {J.}~\bibnamefont {Pixley}},\ }\bibfield  {title}
  {{\color{Gray}\small \bibinfo {title} {Magic-angle semimetals},\ }}\href@noop
  {} {\bibfield  {journal} {\bibinfo  {journal} {arXiv preprint
  arXiv:1809.04604}\ } (\bibinfo {year} {2018})}\BibitemShut {NoStop}%
\bibitem [{\citenamefont {Trambly~de Laissardière}\ \emph
  {et~al.}(2010)\citenamefont {Trambly~de Laissardière}, \citenamefont
  {Mayou},\ and\ \citenamefont {Magaud}}]{doi:10.1021/nl902948m}%
  \BibitemOpen
  \bibfield  {author} {\bibinfo {author} {\bibfnamefont {G.}~\bibnamefont
  {Trambly~de Laissardière}}, \bibinfo {author} {\bibfnamefont
  {D.}~\bibnamefont {Mayou}}, \ and\ \bibinfo {author} {\bibfnamefont
  {L.}~\bibnamefont {Magaud}},\ }\bibfield  {title} {{\color{Gray}\small
  \bibinfo {title} {Localization of dirac electrons in rotated graphene
  bilayers},\ }}\href {\doibase 10.1021/nl902948m} {\bibfield  {journal}
  {\bibinfo  {journal} {Nano Letters}\ }\textbf {\bibinfo {volume} {10}},\
  \bibinfo {pages} {804} (\bibinfo {year} {2010})},\ \bibinfo {note} {pMID:
  20121163},\ \Eprint {http://arxiv.org/abs/https://doi.org/10.1021/nl902948m}
  {https://doi.org/10.1021/nl902948m} \BibitemShut {NoStop}%
\bibitem [{\citenamefont {Bistritzer}\ and\ \citenamefont
  {MacDonald}(2011)}]{Bistritzer12233}%
  \BibitemOpen
  \bibfield  {author} {\bibinfo {author} {\bibfnamefont {R.}~\bibnamefont
  {Bistritzer}}\ and\ \bibinfo {author} {\bibfnamefont {A.~H.}\ \bibnamefont
  {MacDonald}},\ }\bibfield  {title} {{\color{Gray}\small \bibinfo {title}
  {Moir{\'e} bands in twisted double-layer graphene},\ }}\href {\doibase
  10.1073/pnas.1108174108} {\bibfield  {journal} {\bibinfo  {journal}
  {Proceedings of the National Academy of Sciences}\ }\textbf {\bibinfo
  {volume} {108}},\ \bibinfo {pages} {12233} (\bibinfo {year} {2011})},\
  \Eprint
  {http://arxiv.org/abs/https://www.pnas.org/content/108/30/12233.full.pdf}
  {https://www.pnas.org/content/108/30/12233.full.pdf} \BibitemShut {NoStop}%
\bibitem [{\citenamefont {Lopes~dos Santos}\ \emph {et~al.}(2012)\citenamefont
  {Lopes~dos Santos}, \citenamefont {Peres},\ and\ \citenamefont
  {Castro~Neto}}]{PhysRevB.86.155449}%
  \BibitemOpen
  \bibfield  {author} {\bibinfo {author} {\bibfnamefont {J.~M.~B.}\
  \bibnamefont {Lopes~dos Santos}}, \bibinfo {author} {\bibfnamefont
  {N.~M.~R.}\ \bibnamefont {Peres}}, \ and\ \bibinfo {author} {\bibfnamefont
  {A.~H.}\ \bibnamefont {Castro~Neto}},\ }\bibfield  {title}
  {{\color{Gray}\small \bibinfo {title} {Continuum model of the twisted
  graphene bilayer},\ }}\href {\doibase 10.1103/PhysRevB.86.155449} {\bibfield
  {journal} {\bibinfo  {journal} {Phys. Rev. B}\ }\textbf {\bibinfo {volume}
  {86}},\ \bibinfo {pages} {155449} (\bibinfo {year} {2012})}\BibitemShut
  {NoStop}%
\bibitem [{\citenamefont {Li}\ \emph {et~al.}(2010)\citenamefont {Li},
  \citenamefont {Luican}, \citenamefont {Dos~Santos}, \citenamefont {Neto},
  \citenamefont {Reina}, \citenamefont {Kong},\ and\ \citenamefont
  {Andrei}}]{li2010observation}%
  \BibitemOpen
  \bibfield  {author} {\bibinfo {author} {\bibfnamefont {G.}~\bibnamefont
  {Li}}, \bibinfo {author} {\bibfnamefont {A.}~\bibnamefont {Luican}}, \bibinfo
  {author} {\bibfnamefont {J.~L.}\ \bibnamefont {Dos~Santos}}, \bibinfo
  {author} {\bibfnamefont {A.~C.}\ \bibnamefont {Neto}}, \bibinfo {author}
  {\bibfnamefont {A.}~\bibnamefont {Reina}}, \bibinfo {author} {\bibfnamefont
  {J.}~\bibnamefont {Kong}}, \ and\ \bibinfo {author} {\bibfnamefont
  {E.}~\bibnamefont {Andrei}},\ }\bibfield  {title} {{\color{Gray}\small
  \bibinfo {title} {{Observation of Van Hove singularities in twisted graphene
  layers}},\ }}\href@noop {} {\bibfield  {journal} {\bibinfo  {journal} {Nature
  Physics}\ }\textbf {\bibinfo {volume} {6}},\ \bibinfo {pages} {109} (\bibinfo
  {year} {2010})}\BibitemShut {NoStop}%
\bibitem [{\citenamefont {Brihuega}\ \emph {et~al.}(2012)\citenamefont
  {Brihuega}, \citenamefont {Mallet}, \citenamefont {Gonz\'alez-Herrero},
  \citenamefont {Trambly~de Laissardi\`ere}, \citenamefont {Ugeda},
  \citenamefont {Magaud}, \citenamefont {G\'omez-Rodr\'{\i}guez}, \citenamefont
  {Yndur\'ain},\ and\ \citenamefont {Veuillen}}]{PhysRevLett.109.196802}%
  \BibitemOpen
  \bibfield  {author} {\bibinfo {author} {\bibfnamefont {I.}~\bibnamefont
  {Brihuega}}, \bibinfo {author} {\bibfnamefont {P.}~\bibnamefont {Mallet}},
  \bibinfo {author} {\bibfnamefont {H.}~\bibnamefont {Gonz\'alez-Herrero}},
  \bibinfo {author} {\bibfnamefont {G.}~\bibnamefont {Trambly~de
  Laissardi\`ere}}, \bibinfo {author} {\bibfnamefont {M.~M.}\ \bibnamefont
  {Ugeda}}, \bibinfo {author} {\bibfnamefont {L.}~\bibnamefont {Magaud}},
  \bibinfo {author} {\bibfnamefont {J.~M.}\ \bibnamefont
  {G\'omez-Rodr\'{\i}guez}}, \bibinfo {author} {\bibfnamefont {F.}~\bibnamefont
  {Yndur\'ain}}, \ and\ \bibinfo {author} {\bibfnamefont {J.-Y.}\ \bibnamefont
  {Veuillen}},\ }\bibfield  {title} {{\color{Gray}\small \bibinfo {title}
  {{Unraveling the Intrinsic and Robust Nature of van Hove Singularities in
  Twisted Bilayer Graphene by Scanning Tunneling Microscopy and Theoretical
  Analysis}},\ }}\href {\doibase 10.1103/PhysRevLett.109.196802} {\bibfield
  {journal} {\bibinfo  {journal} {Phys. Rev. Lett.}\ }\textbf {\bibinfo
  {volume} {109}},\ \bibinfo {pages} {196802} (\bibinfo {year}
  {2012})}\BibitemShut {NoStop}%
\bibitem [{\citenamefont {Kim}\ \emph {et~al.}(2016)\citenamefont {Kim},
  \citenamefont {Herlinger}, \citenamefont {Moon}, \citenamefont {Koshino},
  \citenamefont {Taniguchi}, \citenamefont {Watanabe},\ and\ \citenamefont
  {Smet}}]{doi:10.1021/acs.nanolett.6b01906}%
  \BibitemOpen
  \bibfield  {author} {\bibinfo {author} {\bibfnamefont {Y.}~\bibnamefont
  {Kim}}, \bibinfo {author} {\bibfnamefont {P.}~\bibnamefont {Herlinger}},
  \bibinfo {author} {\bibfnamefont {P.}~\bibnamefont {Moon}}, \bibinfo {author}
  {\bibfnamefont {M.}~\bibnamefont {Koshino}}, \bibinfo {author} {\bibfnamefont
  {T.}~\bibnamefont {Taniguchi}}, \bibinfo {author} {\bibfnamefont
  {K.}~\bibnamefont {Watanabe}}, \ and\ \bibinfo {author} {\bibfnamefont
  {J.~H.}\ \bibnamefont {Smet}},\ }\bibfield  {title} {{\color{Gray}\small
  \bibinfo {title} {{Charge Inversion and Topological Phase Transition at a
  Twist Angle Induced van Hove Singularity of Bilayer Graphene}},\ }}\href
  {\doibase 10.1021/acs.nanolett.6b01906} {\bibfield  {journal} {\bibinfo
  {journal} {Nano Letters}\ }\textbf {\bibinfo {volume} {16}},\ \bibinfo
  {pages} {5053} (\bibinfo {year} {2016})},\ \bibinfo {note} {pMID: 27387484},\
  \Eprint {http://arxiv.org/abs/https://doi.org/10.1021/acs.nanolett.6b01906}
  {https://doi.org/10.1021/acs.nanolett.6b01906} \BibitemShut {NoStop}%
\bibitem [{\citenamefont {Koshino}\ \emph {et~al.}(2018)\citenamefont
  {Koshino}, \citenamefont {Yuan}, \citenamefont {Koretsune}, \citenamefont
  {Ochi}, \citenamefont {Kuroki},\ and\ \citenamefont
  {Fu}}]{PhysRevX.8.031087}%
  \BibitemOpen
  \bibfield  {author} {\bibinfo {author} {\bibfnamefont {M.}~\bibnamefont
  {Koshino}}, \bibinfo {author} {\bibfnamefont {N.~F.~Q.}\ \bibnamefont
  {Yuan}}, \bibinfo {author} {\bibfnamefont {T.}~\bibnamefont {Koretsune}},
  \bibinfo {author} {\bibfnamefont {M.}~\bibnamefont {Ochi}}, \bibinfo {author}
  {\bibfnamefont {K.}~\bibnamefont {Kuroki}}, \ and\ \bibinfo {author}
  {\bibfnamefont {L.}~\bibnamefont {Fu}},\ }\bibfield  {title}
  {{\color{Gray}\small \bibinfo {title} {{Maximally Localized Wannier Orbitals
  and the Extended Hubbard Model for Twisted Bilayer Graphene}},\ }}\href
  {\doibase 10.1103/PhysRevX.8.031087} {\bibfield  {journal} {\bibinfo
  {journal} {Phys. Rev. X}\ }\textbf {\bibinfo {volume} {8}},\ \bibinfo {pages}
  {031087} (\bibinfo {year} {2018})}\BibitemShut {NoStop}%
\bibitem [{\citenamefont {Kang}\ and\ \citenamefont
  {Vafek}(2018)}]{PhysRevX.8.031088}%
  \BibitemOpen
  \bibfield  {author} {\bibinfo {author} {\bibfnamefont {J.}~\bibnamefont
  {Kang}}\ and\ \bibinfo {author} {\bibfnamefont {O.}~\bibnamefont {Vafek}},\
  }\bibfield  {title} {{\color{Gray}\small \bibinfo {title} {{Symmetry,
  Maximally Localized Wannier States, and a Low-Energy Model for Twisted
  Bilayer Graphene Narrow Bands}},\ }}\href {\doibase
  10.1103/PhysRevX.8.031088} {\bibfield  {journal} {\bibinfo  {journal} {Phys.
  Rev. X}\ }\textbf {\bibinfo {volume} {8}},\ \bibinfo {pages} {031088}
  (\bibinfo {year} {2018})}\BibitemShut {NoStop}%
\bibitem [{\citenamefont {Po}\ \emph {et~al.}(2018)\citenamefont {Po},
  \citenamefont {Zou}, \citenamefont {Vishwanath},\ and\ \citenamefont
  {Senthil}}]{PhysRevX.8.031089}%
  \BibitemOpen
  \bibfield  {author} {\bibinfo {author} {\bibfnamefont {H.~C.}\ \bibnamefont
  {Po}}, \bibinfo {author} {\bibfnamefont {L.}~\bibnamefont {Zou}}, \bibinfo
  {author} {\bibfnamefont {A.}~\bibnamefont {Vishwanath}}, \ and\ \bibinfo
  {author} {\bibfnamefont {T.}~\bibnamefont {Senthil}},\ }\bibfield  {title}
  {{\color{Gray}\small \bibinfo {title} {{Origin of Mott Insulating Behavior
  and Superconductivity in Twisted Bilayer Graphene}},\ }}\href {\doibase
  10.1103/PhysRevX.8.031089} {\bibfield  {journal} {\bibinfo  {journal} {Phys.
  Rev. X}\ }\textbf {\bibinfo {volume} {8}},\ \bibinfo {pages} {031089}
  (\bibinfo {year} {2018})}\BibitemShut {NoStop}%
\bibitem [{\citenamefont {Xu}\ and\ \citenamefont
  {Balents}(2018)}]{PhysRevLett.121.087001}%
  \BibitemOpen
  \bibfield  {author} {\bibinfo {author} {\bibfnamefont {C.}~\bibnamefont
  {Xu}}\ and\ \bibinfo {author} {\bibfnamefont {L.}~\bibnamefont {Balents}},\
  }\bibfield  {title} {{\color{Gray}\small \bibinfo {title} {{Topological
  Superconductivity in Twisted Multilayer Graphene}},\ }}\href {\doibase
  10.1103/PhysRevLett.121.087001} {\bibfield  {journal} {\bibinfo  {journal}
  {Phys. Rev. Lett.}\ }\textbf {\bibinfo {volume} {121}},\ \bibinfo {pages}
  {087001} (\bibinfo {year} {2018})}\BibitemShut {NoStop}%
\bibitem [{\citenamefont {Yuan}\ and\ \citenamefont
  {Fu}(2018)}]{PhysRevB.98.045103}%
  \BibitemOpen
  \bibfield  {author} {\bibinfo {author} {\bibfnamefont {N.~F.~Q.}\
  \bibnamefont {Yuan}}\ and\ \bibinfo {author} {\bibfnamefont {L.}~\bibnamefont
  {Fu}},\ }\bibfield  {title} {{\color{Gray}\small \bibinfo {title} {Model for
  the metal-insulator transition in graphene superlattices and beyond},\
  }}\href {\doibase 10.1103/PhysRevB.98.045103} {\bibfield  {journal} {\bibinfo
   {journal} {Phys. Rev. B}\ }\textbf {\bibinfo {volume} {98}},\ \bibinfo
  {pages} {045103} (\bibinfo {year} {2018})}\BibitemShut {NoStop}%
\bibitem [{\citenamefont {Dodaro}\ \emph {et~al.}(2018)\citenamefont {Dodaro},
  \citenamefont {Kivelson}, \citenamefont {Schattner}, \citenamefont {Sun},\
  and\ \citenamefont {Wang}}]{PhysRevB.98.075154}%
  \BibitemOpen
  \bibfield  {author} {\bibinfo {author} {\bibfnamefont {J.~F.}\ \bibnamefont
  {Dodaro}}, \bibinfo {author} {\bibfnamefont {S.~A.}\ \bibnamefont
  {Kivelson}}, \bibinfo {author} {\bibfnamefont {Y.}~\bibnamefont {Schattner}},
  \bibinfo {author} {\bibfnamefont {X.~Q.}\ \bibnamefont {Sun}}, \ and\
  \bibinfo {author} {\bibfnamefont {C.}~\bibnamefont {Wang}},\ }\bibfield
  {title} {{\color{Gray}\small \bibinfo {title} {Phases of a phenomenological
  model of twisted bilayer graphene},\ }}\href {\doibase
  10.1103/PhysRevB.98.075154} {\bibfield  {journal} {\bibinfo  {journal} {Phys.
  Rev. B}\ }\textbf {\bibinfo {volume} {98}},\ \bibinfo {pages} {075154}
  (\bibinfo {year} {2018})}\BibitemShut {NoStop}%
\bibitem [{\citenamefont {Kennes}\ \emph {et~al.}(2018)\citenamefont {Kennes},
  \citenamefont {Lischner},\ and\ \citenamefont
  {Karrasch}}]{PhysRevB.98.241407}%
  \BibitemOpen
  \bibfield  {author} {\bibinfo {author} {\bibfnamefont {D.~M.}\ \bibnamefont
  {Kennes}}, \bibinfo {author} {\bibfnamefont {J.}~\bibnamefont {Lischner}}, \
  and\ \bibinfo {author} {\bibfnamefont {C.}~\bibnamefont {Karrasch}},\
  }\bibfield  {title} {{\color{Gray}\small \bibinfo {title} {Strong
  correlations and $d+\mathit{id}$ superconductivity in twisted bilayer
  graphene},\ }}\href {\doibase 10.1103/PhysRevB.98.241407} {\bibfield
  {journal} {\bibinfo  {journal} {Phys. Rev. B}\ }\textbf {\bibinfo {volume}
  {98}},\ \bibinfo {pages} {241407} (\bibinfo {year} {2018})}\BibitemShut
  {NoStop}%
\bibitem [{\citenamefont {Tang}\ \emph {et~al.}(2018)\citenamefont {Tang},
  \citenamefont {Yang}, \citenamefont {Wang}, \citenamefont {Zhang},\ and\
  \citenamefont {Wang}}]{tang2018spin}%
  \BibitemOpen
  \bibfield  {author} {\bibinfo {author} {\bibfnamefont {Q.}~\bibnamefont
  {Tang}}, \bibinfo {author} {\bibfnamefont {L.}~\bibnamefont {Yang}}, \bibinfo
  {author} {\bibfnamefont {D.}~\bibnamefont {Wang}}, \bibinfo {author}
  {\bibfnamefont {F.}~\bibnamefont {Zhang}}, \ and\ \bibinfo {author}
  {\bibfnamefont {Q.}~\bibnamefont {Wang}},\ }\bibfield  {title}
  {{\color{Gray}\small \bibinfo {title} {Spin-triplet $ f $-wave pairing in
  twisted bilayer graphene near 1/4 filling},\ }}\href@noop {} {\bibfield
  {journal} {\bibinfo  {journal} {arXiv preprint arXiv:1809.06772}\ } (\bibinfo
  {year} {2018})}\BibitemShut {NoStop}%
\bibitem [{\citenamefont {Isobe}\ \emph {et~al.}(2018)\citenamefont {Isobe},
  \citenamefont {Yuan},\ and\ \citenamefont {Fu}}]{PhysRevX.8.041041}%
  \BibitemOpen
  \bibfield  {author} {\bibinfo {author} {\bibfnamefont {H.}~\bibnamefont
  {Isobe}}, \bibinfo {author} {\bibfnamefont {N.~F.~Q.}\ \bibnamefont {Yuan}},
  \ and\ \bibinfo {author} {\bibfnamefont {L.}~\bibnamefont {Fu}},\ }\bibfield
  {title} {{\color{Gray}\small \bibinfo {title} {Unconventional
  superconductivity and density waves in twisted bilayer graphene},\ }}\href
  {\doibase 10.1103/PhysRevX.8.041041} {\bibfield  {journal} {\bibinfo
  {journal} {Phys. Rev. X}\ }\textbf {\bibinfo {volume} {8}},\ \bibinfo {pages}
  {041041} (\bibinfo {year} {2018})}\BibitemShut {NoStop}%
\bibitem [{\citenamefont {Lin}\ and\ \citenamefont
  {Nandkishore}(2019)}]{lin2019chiral}%
  \BibitemOpen
  \bibfield  {author} {\bibinfo {author} {\bibfnamefont {Y.-P.}\ \bibnamefont
  {Lin}}\ and\ \bibinfo {author} {\bibfnamefont {R.~M.}\ \bibnamefont
  {Nandkishore}},\ }\bibfield  {title} {{\color{Gray}\small \bibinfo {title}
  {{A chiral twist on the high-$ T\_c $ phase diagram in Moir\'e
  heterostructures}},\ }}\href@noop {} {\bibfield  {journal} {\bibinfo
  {journal} {arXiv preprint arXiv:1901.00500}\ } (\bibinfo {year}
  {2019})}\BibitemShut {NoStop}%
\bibitem [{\citenamefont {Rademaker}\ and\ \citenamefont
  {Mellado}(2018)}]{PhysRevB.98.235158}%
  \BibitemOpen
  \bibfield  {author} {\bibinfo {author} {\bibfnamefont {L.}~\bibnamefont
  {Rademaker}}\ and\ \bibinfo {author} {\bibfnamefont {P.}~\bibnamefont
  {Mellado}},\ }\bibfield  {title} {{\color{Gray}\small \bibinfo {title}
  {Charge-transfer insulation in twisted bilayer graphene},\ }}\href {\doibase
  10.1103/PhysRevB.98.235158} {\bibfield  {journal} {\bibinfo  {journal} {Phys.
  Rev. B}\ }\textbf {\bibinfo {volume} {98}},\ \bibinfo {pages} {235158}
  (\bibinfo {year} {2018})}\BibitemShut {NoStop}%
\bibitem [{\citenamefont {Huang}\ \emph {et~al.}(2018)\citenamefont {Huang},
  \citenamefont {Zhang},\ and\ \citenamefont
  {Ma}}]{huang2018antiferromagnetically}%
  \BibitemOpen
  \bibfield  {author} {\bibinfo {author} {\bibfnamefont {T.}~\bibnamefont
  {Huang}}, \bibinfo {author} {\bibfnamefont {L.}~\bibnamefont {Zhang}}, \ and\
  \bibinfo {author} {\bibfnamefont {T.}~\bibnamefont {Ma}},\ }\bibfield
  {title} {{\color{Gray}\small \bibinfo {title} {{Antiferromagnetically ordered
  Mott insulator and $ d+ id $ superconductivity in twisted bilayer graphene: A
  quantum Monte carlo study}},\ }}\href@noop {} {\bibfield  {journal} {\bibinfo
   {journal} {arXiv preprint arXiv:1804.06096}\ } (\bibinfo {year}
  {2018})}\BibitemShut {NoStop}%
\bibitem [{\citenamefont {Fidrysiak}\ \emph {et~al.}(2018)\citenamefont
  {Fidrysiak}, \citenamefont {Zegrodnik},\ and\ \citenamefont
  {Spa\l{}ek}}]{PhysRevB.98.085436}%
  \BibitemOpen
  \bibfield  {author} {\bibinfo {author} {\bibfnamefont {M.}~\bibnamefont
  {Fidrysiak}}, \bibinfo {author} {\bibfnamefont {M.}~\bibnamefont
  {Zegrodnik}}, \ and\ \bibinfo {author} {\bibfnamefont {J.}~\bibnamefont
  {Spa\l{}ek}},\ }\bibfield  {title} {{\color{Gray}\small \bibinfo {title}
  {Unconventional topological superconductivity and phase diagram for an
  effective two-orbital model as applied to twisted bilayer graphene},\ }}\href
  {\doibase 10.1103/PhysRevB.98.085436} {\bibfield  {journal} {\bibinfo
  {journal} {Phys. Rev. B}\ }\textbf {\bibinfo {volume} {98}},\ \bibinfo
  {pages} {085436} (\bibinfo {year} {2018})}\BibitemShut {NoStop}%
\bibitem [{\citenamefont {Guo}\ \emph {et~al.}(2018)\citenamefont {Guo},
  \citenamefont {Zhu}, \citenamefont {Feng},\ and\ \citenamefont
  {Scalettar}}]{PhysRevB.97.235453}%
  \BibitemOpen
  \bibfield  {author} {\bibinfo {author} {\bibfnamefont {H.}~\bibnamefont
  {Guo}}, \bibinfo {author} {\bibfnamefont {X.}~\bibnamefont {Zhu}}, \bibinfo
  {author} {\bibfnamefont {S.}~\bibnamefont {Feng}}, \ and\ \bibinfo {author}
  {\bibfnamefont {R.~T.}\ \bibnamefont {Scalettar}},\ }\bibfield  {title}
  {{\color{Gray}\small \bibinfo {title} {Pairing symmetry of interacting
  fermions on a twisted bilayer graphene superlattice},\ }}\href {\doibase
  10.1103/PhysRevB.97.235453} {\bibfield  {journal} {\bibinfo  {journal} {Phys.
  Rev. B}\ }\textbf {\bibinfo {volume} {97}},\ \bibinfo {pages} {235453}
  (\bibinfo {year} {2018})}\BibitemShut {NoStop}%
\bibitem [{\citenamefont {Roy}\ and\ \citenamefont
  {Juricic}(2018)}]{roy2018unconventional}%
  \BibitemOpen
  \bibfield  {author} {\bibinfo {author} {\bibfnamefont {B.}~\bibnamefont
  {Roy}}\ and\ \bibinfo {author} {\bibfnamefont {V.}~\bibnamefont {Juricic}},\
  }\bibfield  {title} {{\color{Gray}\small \bibinfo {title} {Unconventional
  superconductivity in nearly flat bands in twisted bilayer graphene},\
  }}\href@noop {} {\bibfield  {journal} {\bibinfo  {journal} {arXiv preprint
  arXiv:1803.11190}\ } (\bibinfo {year} {2018})}\BibitemShut {NoStop}%
\bibitem [{\citenamefont {Liu}\ \emph {et~al.}(2018)\citenamefont {Liu},
  \citenamefont {Zhang}, \citenamefont {Chen},\ and\ \citenamefont
  {Yang}}]{PhysRevLett.121.217001}%
  \BibitemOpen
  \bibfield  {author} {\bibinfo {author} {\bibfnamefont {C.-C.}\ \bibnamefont
  {Liu}}, \bibinfo {author} {\bibfnamefont {L.-D.}\ \bibnamefont {Zhang}},
  \bibinfo {author} {\bibfnamefont {W.-Q.}\ \bibnamefont {Chen}}, \ and\
  \bibinfo {author} {\bibfnamefont {F.}~\bibnamefont {Yang}},\ }\bibfield
  {title} {{\color{Gray}\small \bibinfo {title} {Chiral spin density wave and
  $d+id$ superconductivity in the magic-angle-twisted bilayer graphene},\
  }}\href {\doibase 10.1103/PhysRevLett.121.217001} {\bibfield  {journal}
  {\bibinfo  {journal} {Phys. Rev. Lett.}\ }\textbf {\bibinfo {volume} {121}},\
  \bibinfo {pages} {217001} (\bibinfo {year} {2018})}\BibitemShut {NoStop}%
\bibitem [{\citenamefont {Da~Liao}\ \emph {et~al.}(2019)\citenamefont
  {Da~Liao}, \citenamefont {Meng},\ and\ \citenamefont {Xu}}]{da2019magic}%
  \BibitemOpen
  \bibfield  {author} {\bibinfo {author} {\bibfnamefont {Y.}~\bibnamefont
  {Da~Liao}}, \bibinfo {author} {\bibfnamefont {Z.~Y.}\ \bibnamefont {Meng}}, \
  and\ \bibinfo {author} {\bibfnamefont {X.~Y.}\ \bibnamefont {Xu}},\
  }\bibfield  {title} {{\color{Gray}\small \bibinfo {title} {Is magic-angle
  twisted bilayer graphene near a quantum critical point?}\ }}\href@noop {}
  {\bibfield  {journal} {\bibinfo  {journal} {arXiv preprint arXiv:1901.11424}\
  } (\bibinfo {year} {2019})}\BibitemShut {NoStop}%
\bibitem [{\citenamefont {{Xian}}\ \emph {et~al.}(2018)\citenamefont {{Xian}},
  \citenamefont {{Kennes}}, \citenamefont {{Tancogne-Dejean}}, \citenamefont
  {{Altarelli}},\ and\ \citenamefont {{Rubio}}}]{2018arXiv181208097X}%
  \BibitemOpen
  \bibfield  {author} {\bibinfo {author} {\bibfnamefont {L.}~\bibnamefont
  {{Xian}}}, \bibinfo {author} {\bibfnamefont {D.~M.}\ \bibnamefont
  {{Kennes}}}, \bibinfo {author} {\bibfnamefont {N.}~\bibnamefont
  {{Tancogne-Dejean}}}, \bibinfo {author} {\bibfnamefont {M.}~\bibnamefont
  {{Altarelli}}}, \ and\ \bibinfo {author} {\bibfnamefont {A.}~\bibnamefont
  {{Rubio}}},\ }\bibfield  {title} {{\color{Gray}\small \bibinfo {title}
  {{Multi-flat bands and strong correlations in Twisted Bilayer Boron
  Nitride}},\ }}\href@noop {} {\bibfield  {journal} {\bibinfo  {journal} {arXiv
  e-prints}\ ,\ \bibinfo {eid} {arXiv:1812.08097}} (\bibinfo {year} {2018})},\
  \Eprint {http://arxiv.org/abs/1812.08097} {arXiv:1812.08097
  [cond-mat.mes-hall]} \BibitemShut {NoStop}%
\bibitem [{\citenamefont {Chittari}\ \emph {et~al.}(2019)\citenamefont
  {Chittari}, \citenamefont {Chen}, \citenamefont {Zhang}, \citenamefont
  {Wang},\ and\ \citenamefont {Jung}}]{PhysRevLett.122.016401}%
  \BibitemOpen
  \bibfield  {author} {\bibinfo {author} {\bibfnamefont {B.~L.}\ \bibnamefont
  {Chittari}}, \bibinfo {author} {\bibfnamefont {G.}~\bibnamefont {Chen}},
  \bibinfo {author} {\bibfnamefont {Y.}~\bibnamefont {Zhang}}, \bibinfo
  {author} {\bibfnamefont {F.}~\bibnamefont {Wang}}, \ and\ \bibinfo {author}
  {\bibfnamefont {J.}~\bibnamefont {Jung}},\ }\bibfield  {title}
  {{\color{Gray}\small \bibinfo {title} {Gate-tunable topological flat bands in
  trilayer graphene boron-nitride moir\'e superlattices},\ }}\href {\doibase
  10.1103/PhysRevLett.122.016401} {\bibfield  {journal} {\bibinfo  {journal}
  {Phys. Rev. Lett.}\ }\textbf {\bibinfo {volume} {122}},\ \bibinfo {pages}
  {016401} (\bibinfo {year} {2019})}\BibitemShut {NoStop}%
\bibitem [{\citenamefont {Metzner}\ \emph {et~al.}(2012)\citenamefont
  {Metzner}, \citenamefont {Salmhofer}, \citenamefont {Honerkamp},
  \citenamefont {Meden},\ and\ \citenamefont
  {Sch\"onhammer}}]{RevModPhys.84.299}%
  \BibitemOpen
  \bibfield  {author} {\bibinfo {author} {\bibfnamefont {W.}~\bibnamefont
  {Metzner}}, \bibinfo {author} {\bibfnamefont {M.}~\bibnamefont {Salmhofer}},
  \bibinfo {author} {\bibfnamefont {C.}~\bibnamefont {Honerkamp}}, \bibinfo
  {author} {\bibfnamefont {V.}~\bibnamefont {Meden}}, \ and\ \bibinfo {author}
  {\bibfnamefont {K.}~\bibnamefont {Sch\"onhammer}},\ }\bibfield  {title}
  {{\color{Gray}\small \bibinfo {title} {{Functional renormalization group
  approach to correlated fermion systems}},\ }}\href {\doibase
  10.1103/RevModPhys.84.299} {\bibfield  {journal} {\bibinfo  {journal} {Rev.
  Mod. Phys.}\ }\textbf {\bibinfo {volume} {84}},\ \bibinfo {pages} {299}
  (\bibinfo {year} {2012})}\BibitemShut {NoStop}%
\bibitem [{\citenamefont {Platt}\ \emph {et~al.}(2013)\citenamefont {Platt},
  \citenamefont {Hanke},\ and\ \citenamefont
  {Thomale}}]{doi:10.1080/00018732.2013.862020}%
  \BibitemOpen
  \bibfield  {author} {\bibinfo {author} {\bibfnamefont {C.}~\bibnamefont
  {Platt}}, \bibinfo {author} {\bibfnamefont {W.}~\bibnamefont {Hanke}}, \ and\
  \bibinfo {author} {\bibfnamefont {R.}~\bibnamefont {Thomale}},\ }\bibfield
  {title} {{\color{Gray}\small \bibinfo {title} {{Functional renormalization
  group for multi-orbital Fermi surface instabilities}},\ }}\href {\doibase
  10.1080/00018732.2013.862020} {\bibfield  {journal} {\bibinfo  {journal}
  {Advances in Physics}\ }\textbf {\bibinfo {volume} {62}},\ \bibinfo {pages}
  {453} (\bibinfo {year} {2013})}\BibitemShut {NoStop}%
\bibitem [{\citenamefont {Wetterich}(1993)}]{Wetterich:1992yh}%
  \BibitemOpen
  \bibfield  {author} {\bibinfo {author} {\bibfnamefont {C.}~\bibnamefont
  {Wetterich}},\ }\bibfield  {title} {{\color{Gray}\small \bibinfo {title}
  {{Exact evolution equation for the effective potential}},\ }}\href {\doibase
  10.1016/0370-2693(93)90726-X} {\bibfield  {journal} {\bibinfo  {journal}
  {Phys. Lett.}\ }\textbf {\bibinfo {volume} {B301}},\ \bibinfo {pages} {90}
  (\bibinfo {year} {1993})},\ \Eprint {http://arxiv.org/abs/1710.05815}
  {arXiv:1710.05815 [hep-th]} \BibitemShut {NoStop}%
\bibitem [{\citenamefont {Venderbos}(2016)}]{PhysRevB.93.115107}%
  \BibitemOpen
  \bibfield  {author} {\bibinfo {author} {\bibfnamefont {J.~W.~F.}\
  \bibnamefont {Venderbos}},\ }\bibfield  {title} {{\color{Gray}\small \bibinfo
  {title} {Symmetry analysis of translational symmetry broken density waves:
  Application to hexagonal lattices in two dimensions},\ }}\href {\doibase
  10.1103/PhysRevB.93.115107} {\bibfield  {journal} {\bibinfo  {journal} {Phys.
  Rev. B}\ }\textbf {\bibinfo {volume} {93}},\ \bibinfo {pages} {115107}
  (\bibinfo {year} {2016})}\BibitemShut {NoStop}%
\bibitem [{Note1()}]{Note1}%
  \BibitemOpen
  \bibinfo {note} {The couplings in Eqs.~\protect \textup {\hbox {\mathsurround
  \z@ \protect \normalfont (\ignorespaces \ref {eq:ia0}\unskip \@@italiccorr
  )}}, \protect \textup {\hbox {\mathsurround \z@ \protect \normalfont
  (\ignorespaces \ref {eq:ia02}\unskip \@@italiccorr )}} relate to the common
  spin-SU(2) symmetric Hubbard and Hund terms $H_I=1/2\DOTSB \sum@ \slimits@
  _{\sigma \sigma '}[\DOTSB \sum@ \slimits@ _{o}U_{\protect \mathrm {SU(2)}}
  n_{io\sigma }n_{io\sigma '} + U'_{\protect \mathrm {SU(2)}} \DOTSB \sum@
  \slimits@ _{o\not =o'} n_{io\sigma }n_{io'\sigma '}+J_{\protect \mathrm
  {SU(2)}}\DOTSB \sum@ \slimits@ _{o\not =o'}c^\dagger _{io\sigma }c^\dagger
  _{io'\sigma '}c_{io\sigma '}c_{io'\sigma '}]$ via $U_{\protect \mathrm
  {SU(2)}}=U+V/2+K/2$, $U'_{\protect \mathrm {SU(2)}}=U_{\protect \mathrm
  {SU(2)}}$ and $J_{\protect \mathrm {SU(2)}}=V-K$. Pair-hopping $J_{\protect
  \mathrm {SU(2)}}'$ and $U_{\protect \mathrm {SU(2)}}\not =U'_{\protect
  \mathrm {SU(2)}}$ would break the SU(2)$\times $SU(2) symmetry. The term
  $\propto V$ can be absorbed by a shift into the terms $\propto U, K$~\cite
  {PhysRevB.98.075154}: $U\rightarrow U-V$, $K\rightarrow K+V$.}\BibitemShut
  {Stop}%
\bibitem [{\citenamefont {Nandkishore}\ \emph
  {et~al.}(2012{\natexlab{a}})\citenamefont {Nandkishore}, \citenamefont
  {Chern},\ and\ \citenamefont {Chubukov}}]{PhysRevLett.108.227204}%
  \BibitemOpen
  \bibfield  {author} {\bibinfo {author} {\bibfnamefont {R.}~\bibnamefont
  {Nandkishore}}, \bibinfo {author} {\bibfnamefont {G.-W.}\ \bibnamefont
  {Chern}}, \ and\ \bibinfo {author} {\bibfnamefont {A.~V.}\ \bibnamefont
  {Chubukov}},\ }\bibfield  {title} {{\color{Gray}\small \bibinfo {title}
  {Itinerant half-metal spin-density-wave state on the hexagonal lattice},\
  }}\href {\doibase 10.1103/PhysRevLett.108.227204} {\bibfield  {journal}
  {\bibinfo  {journal} {Phys. Rev. Lett.}\ }\textbf {\bibinfo {volume} {108}},\
  \bibinfo {pages} {227204} (\bibinfo {year} {2012}{\natexlab{a}})}\BibitemShut
  {NoStop}%
\bibitem [{\citenamefont {Li}(2012)}]{0295-5075-97-3-37001}%
  \BibitemOpen
  \bibfield  {author} {\bibinfo {author} {\bibfnamefont {T.}~\bibnamefont
  {Li}},\ }\bibfield  {title} {{\color{Gray}\small \bibinfo {title}
  {Spontaneous quantum hall effect in quarter-doped hubbard model on honeycomb
  lattice and its possible realization in doped graphene system},\ }}\href
  {http://stacks.iop.org/0295-5075/97/i=3/a=37001} {\bibfield  {journal}
  {\bibinfo  {journal} {EPL (Europhysics Letters)}\ }\textbf {\bibinfo {volume}
  {97}},\ \bibinfo {pages} {37001} (\bibinfo {year} {2012})}\BibitemShut
  {NoStop}%
\bibitem [{\citenamefont {Wang}\ \emph {et~al.}(2012)\citenamefont {Wang},
  \citenamefont {Xiang}, \citenamefont {Wang}, \citenamefont {Wang},
  \citenamefont {Yang},\ and\ \citenamefont {Lee}}]{PhysRevB.85.035414}%
  \BibitemOpen
  \bibfield  {author} {\bibinfo {author} {\bibfnamefont {W.-S.}\ \bibnamefont
  {Wang}}, \bibinfo {author} {\bibfnamefont {Y.-Y.}\ \bibnamefont {Xiang}},
  \bibinfo {author} {\bibfnamefont {Q.-H.}\ \bibnamefont {Wang}}, \bibinfo
  {author} {\bibfnamefont {F.}~\bibnamefont {Wang}}, \bibinfo {author}
  {\bibfnamefont {F.}~\bibnamefont {Yang}}, \ and\ \bibinfo {author}
  {\bibfnamefont {D.-H.}\ \bibnamefont {Lee}},\ }\bibfield  {title}
  {{\color{Gray}\small \bibinfo {title} {Functional renormalization group and
  variational monte carlo studies of the electronic instabilities in graphene
  near $\frac{1}{4}$ doping},\ }}\href {\doibase 10.1103/PhysRevB.85.035414}
  {\bibfield  {journal} {\bibinfo  {journal} {Phys. Rev. B}\ }\textbf {\bibinfo
  {volume} {85}},\ \bibinfo {pages} {035414} (\bibinfo {year}
  {2012})}\BibitemShut {NoStop}%
\bibitem [{\citenamefont {Martin}\ and\ \citenamefont
  {Batista}(2008)}]{PhysRevLett.101.156402}%
  \BibitemOpen
  \bibfield  {author} {\bibinfo {author} {\bibfnamefont {I.}~\bibnamefont
  {Martin}}\ and\ \bibinfo {author} {\bibfnamefont {C.~D.}\ \bibnamefont
  {Batista}},\ }\bibfield  {title} {{\color{Gray}\small \bibinfo {title}
  {Itinerant electron-driven chiral magnetic ordering and spontaneous quantum
  hall effect in triangular lattice models},\ }}\href {\doibase
  10.1103/PhysRevLett.101.156402} {\bibfield  {journal} {\bibinfo  {journal}
  {Phys. Rev. Lett.}\ }\textbf {\bibinfo {volume} {101}},\ \bibinfo {pages}
  {156402} (\bibinfo {year} {2008})}\BibitemShut {NoStop}%
\bibitem [{\citenamefont {Black-Schaffer}\ and\ \citenamefont
  {Honerkamp}(2014)}]{0953-8984-26-42-423201}%
  \BibitemOpen
  \bibfield  {author} {\bibinfo {author} {\bibfnamefont {A.~M.}\ \bibnamefont
  {Black-Schaffer}}\ and\ \bibinfo {author} {\bibfnamefont {C.}~\bibnamefont
  {Honerkamp}},\ }\bibfield  {title} {{\color{Gray}\small \bibinfo {title}
  {Chiral $d$-wave superconductivity in doped graphene},\ }}\href
  {http://stacks.iop.org/0953-8984/26/i=42/a=423201} {\bibfield  {journal}
  {\bibinfo  {journal} {Journal of Physics: Condensed Matter}\ }\textbf
  {\bibinfo {volume} {26}},\ \bibinfo {pages} {423201} (\bibinfo {year}
  {2014})}\BibitemShut {NoStop}%
\bibitem [{\citenamefont {Honerkamp}(2003)}]{PhysRevB.68.104510}%
  \BibitemOpen
  \bibfield  {author} {\bibinfo {author} {\bibfnamefont {C.}~\bibnamefont
  {Honerkamp}},\ }\bibfield  {title} {{\color{Gray}\small \bibinfo {title}
  {Instabilities of interacting electrons on the triangular lattice},\ }}\href
  {\doibase 10.1103/PhysRevB.68.104510} {\bibfield  {journal} {\bibinfo
  {journal} {Phys. Rev. B}\ }\textbf {\bibinfo {volume} {68}},\ \bibinfo
  {pages} {104510} (\bibinfo {year} {2003})}\BibitemShut {NoStop}%
\bibitem [{\citenamefont {Nandkishore}\ \emph
  {et~al.}(2012{\natexlab{b}})\citenamefont {Nandkishore}, \citenamefont
  {Levitov},\ and\ \citenamefont {Chubukov}}]{nandkishore2012chiral}%
  \BibitemOpen
  \bibfield  {author} {\bibinfo {author} {\bibfnamefont {R.}~\bibnamefont
  {Nandkishore}}, \bibinfo {author} {\bibfnamefont {L.}~\bibnamefont
  {Levitov}}, \ and\ \bibinfo {author} {\bibfnamefont {A.}~\bibnamefont
  {Chubukov}},\ }\bibfield  {title} {{\color{Gray}\small \bibinfo {title}
  {Chiral superconductivity from repulsive interactions in doped graphene},\
  }}\href@noop {} {\bibfield  {journal} {\bibinfo  {journal} {Nature Physics}\
  }\textbf {\bibinfo {volume} {8}},\ \bibinfo {pages} {158} (\bibinfo {year}
  {2012}{\natexlab{b}})}\BibitemShut {NoStop}%
\bibitem [{\citenamefont {Venderbos}\ and\ \citenamefont
  {Fernandes}(2018)}]{PhysRevB.98.245103}%
  \BibitemOpen
  \bibfield  {author} {\bibinfo {author} {\bibfnamefont {J.~W.~F.}\
  \bibnamefont {Venderbos}}\ and\ \bibinfo {author} {\bibfnamefont {R.~M.}\
  \bibnamefont {Fernandes}},\ }\bibfield  {title} {{\color{Gray}\small \bibinfo
  {title} {Correlations and electronic order in a two-orbital honeycomb lattice
  model for twisted bilayer graphene},\ }}\href {\doibase
  10.1103/PhysRevB.98.245103} {\bibfield  {journal} {\bibinfo  {journal} {Phys.
  Rev. B}\ }\textbf {\bibinfo {volume} {98}},\ \bibinfo {pages} {245103}
  (\bibinfo {year} {2018})}\BibitemShut {NoStop}%
\bibitem [{\citenamefont {Sch\"uler}\ \emph {et~al.}(2013)\citenamefont
  {Sch\"uler}, \citenamefont {R\"osner}, \citenamefont {Wehling}, \citenamefont
  {Lichtenstein},\ and\ \citenamefont {Katsnelson}}]{PhysRevLett.111.036601}%
  \BibitemOpen
  \bibfield  {author} {\bibinfo {author} {\bibfnamefont {M.}~\bibnamefont
  {Sch\"uler}}, \bibinfo {author} {\bibfnamefont {M.}~\bibnamefont {R\"osner}},
  \bibinfo {author} {\bibfnamefont {T.~O.}\ \bibnamefont {Wehling}}, \bibinfo
  {author} {\bibfnamefont {A.~I.}\ \bibnamefont {Lichtenstein}}, \ and\
  \bibinfo {author} {\bibfnamefont {M.~I.}\ \bibnamefont {Katsnelson}},\
  }\bibfield  {title} {{\color{Gray}\small \bibinfo {title} {Optimal hubbard
  models for materials with nonlocal coulomb interactions: Graphene, silicene,
  and benzene},\ }}\href {\doibase 10.1103/PhysRevLett.111.036601} {\bibfield
  {journal} {\bibinfo  {journal} {Phys. Rev. Lett.}\ }\textbf {\bibinfo
  {volume} {111}},\ \bibinfo {pages} {036601} (\bibinfo {year}
  {2013})}\BibitemShut {NoStop}%
\bibitem [{\citenamefont {Honerkamp}\ \emph {et~al.}(2001)\citenamefont
  {Honerkamp}, \citenamefont {Salmhofer}, \citenamefont {Furukawa},\ and\
  \citenamefont {Rice}}]{PhysRevB.63.035109}%
  \BibitemOpen
  \bibfield  {author} {\bibinfo {author} {\bibfnamefont {C.}~\bibnamefont
  {Honerkamp}}, \bibinfo {author} {\bibfnamefont {M.}~\bibnamefont
  {Salmhofer}}, \bibinfo {author} {\bibfnamefont {N.}~\bibnamefont {Furukawa}},
  \ and\ \bibinfo {author} {\bibfnamefont {T.~M.}\ \bibnamefont {Rice}},\
  }\bibfield  {title} {{\color{Gray}\small \bibinfo {title} {Breakdown of the
  landau-fermi liquid in two dimensions due to umklapp scattering},\ }}\href
  {\doibase 10.1103/PhysRevB.63.035109} {\bibfield  {journal} {\bibinfo
  {journal} {Phys. Rev. B}\ }\textbf {\bibinfo {volume} {63}},\ \bibinfo
  {pages} {035109} (\bibinfo {year} {2001})}\BibitemShut {NoStop}%
\bibitem [{\citenamefont {Salmhofer}\ and\ \citenamefont
  {Honerkamp}(2001)}]{doi:10.1143/PTP.105.1}%
  \BibitemOpen
  \bibfield  {author} {\bibinfo {author} {\bibfnamefont {M.}~\bibnamefont
  {Salmhofer}}\ and\ \bibinfo {author} {\bibfnamefont {C.}~\bibnamefont
  {Honerkamp}},\ }\bibfield  {title} {{\color{Gray}\small \bibinfo {title}
  {{Fermionic Renormalization Group Flows: Technique and Theory}},\ }}\href
  {\doibase 10.1143/PTP.105.1} {\bibfield  {journal} {\bibinfo  {journal}
  {Progress of Theoretical Physics}\ }\textbf {\bibinfo {volume} {105}},\
  \bibinfo {pages} {1} (\bibinfo {year} {2001})}\BibitemShut {NoStop}%
\bibitem [{\citenamefont {Hesselmann}\ \emph {et~al.}(2018)\citenamefont
  {Hesselmann}, \citenamefont {Scherer}, \citenamefont {Scherer},\ and\
  \citenamefont {Wessel}}]{PhysRevB.98.045142}%
  \BibitemOpen
  \bibfield  {author} {\bibinfo {author} {\bibfnamefont {S.}~\bibnamefont
  {Hesselmann}}, \bibinfo {author} {\bibfnamefont {D.~D.}\ \bibnamefont
  {Scherer}}, \bibinfo {author} {\bibfnamefont {M.~M.}\ \bibnamefont
  {Scherer}}, \ and\ \bibinfo {author} {\bibfnamefont {S.}~\bibnamefont
  {Wessel}},\ }\bibfield  {title} {{\color{Gray}\small \bibinfo {title}
  {Bond-ordered states and $f$-wave pairing of spinless fermions on the
  honeycomb lattice},\ }}\href {\doibase 10.1103/PhysRevB.98.045142} {\bibfield
   {journal} {\bibinfo  {journal} {Phys. Rev. B}\ }\textbf {\bibinfo {volume}
  {98}},\ \bibinfo {pages} {045142} (\bibinfo {year} {2018})}\BibitemShut
  {NoStop}%
\bibitem [{\citenamefont {Scherer}\ \emph {et~al.}(2014)\citenamefont
  {Scherer}, \citenamefont {Scherer}, \citenamefont {Khaliullin}, \citenamefont
  {Honerkamp},\ and\ \citenamefont {Rosenow}}]{PhysRevB.90.045135}%
  \BibitemOpen
  \bibfield  {author} {\bibinfo {author} {\bibfnamefont {D.~D.}\ \bibnamefont
  {Scherer}}, \bibinfo {author} {\bibfnamefont {M.~M.}\ \bibnamefont
  {Scherer}}, \bibinfo {author} {\bibfnamefont {G.}~\bibnamefont {Khaliullin}},
  \bibinfo {author} {\bibfnamefont {C.}~\bibnamefont {Honerkamp}}, \ and\
  \bibinfo {author} {\bibfnamefont {B.}~\bibnamefont {Rosenow}},\ }\bibfield
  {title} {{\color{Gray}\small \bibinfo {title} {{Unconventional pairing and
  electronic dimerization instabilities in the doped Kitaev-Heisenberg
  model}},\ }}\href {\doibase 10.1103/PhysRevB.90.045135} {\bibfield  {journal}
  {\bibinfo  {journal} {Phys. Rev. B}\ }\textbf {\bibinfo {volume} {90}},\
  \bibinfo {pages} {045135} (\bibinfo {year} {2014})}\BibitemShut {NoStop}%
\end{thebibliography}%

\end{document}